\documentclass[pre,showkeys,superscriptaddress,twocolumn]{revtex4}
\pdfoutput=1
\usepackage[colorlinks=true,urlcolor=blue]{hyperref}
\usepackage{amsfonts}
\usepackage{graphicx}
\usepackage{amsmath}
\usepackage{xcolor}
\usepackage{bm}

\usepackage{color}
\definecolor{red}{rgb}{0.75,0,0}
\definecolor{blue}{rgb}{0,0,0.75}
\definecolor{green}{rgb}{0,0.5,0}

\newcommand{\arxivhref}[2]{{\hypersetup{urlcolor=green}\href{#1}{#2}}}

\DeclareMathOperator{\tr}{tr}

\begin{document}

\title{Defect dynamics in active nematics}

\author{Luca Giomi}
\affiliation{SISSA, International School for Advanced Studies, Via Bonomea 265, 34136 Trieste, Italy}
\author{Mark J. Bowick}
\affiliation{Physics Department and Syracuse Biomaterials Institute, Syracuse University, Syracuse NY 13244, USA}
\author{Prashant Mishra}
\affiliation{Physics Department, Syracuse University, Syracuse NY 13244, USA}
\author{Rastko Sknepnek}
\affiliation{School of Engineering, Physics, and Mathematics, University of Dundee, Dundee DD1 4HN, UK}
\author{M.~Cristina Marchetti}
\affiliation{Physics Department and Syracuse Biomaterials Institute, Syracuse University, Syracuse NY 13244, USA}

\begin{abstract}
Topological defects are distinctive signatures of liquid crystals. They profoundly affect the viscoelastic behavior  of the fluid by constraining the orientational structure  in a way that inevitably requires global changes not achievable with any set of local  deformations. In active nematic liquid crystals topological defects  not only dictate the global structure of the director, but also act as local sources of motion, behaving as self-propelled particles.	In this article we present a detailed analytical and numerical study of the mechanics of topological defects in active nematic liquid crystals.
\end{abstract}


\maketitle

\section{Introduction}

Active liquid crystals are nonequilibrium fluids composed of internally driven elongated units. Examples of active systems that can exhibit liquid crystalline order include mixtures of cytoskeletal filaments and associated motor proteins, bacterial suspensions, the cell cytoskeleton and even non-living analogues, such as monolayers of vibrated granular rods \cite{Marchetti:2013}. The key feature that distinguishes active liquid crystals from their well-studied passive counterparts is that they are maintained out of equilibrium not by an external force applied at the system's boundary, such as an imposed shear, but by an energy input on each individual unit. The energy fed into the system at the microscopic scale is then transformed into organized motion at the large scale. This type of ``reverse energy cascade'' is the hallmark of active systems. In active liquid crystals the large scale self-organized flows resulting from activity further couple to orientational order,  yielding a very rich behavior.   Novel effects that have been predicted theoretically or observed in simulations and experiments include  spontaneous laminar flow \cite{Voituriez:2005,Marenduzzo:2007,Giomi:2008}, large density fluctuations \cite{Ramaswamy:2003,Mishra:2006,Narayan:2007}, unusual rheological properties \cite{Sokolov:2009,Giomi:2010,Fielding:2011}, excitability \cite{Giomi:2011,Giomi:2012} and low Reynolds number ``turbulence'' \cite{Giomi:2011,Giomi:2012,Wensink:2012,Thampi:2013,Thampi:2014a,Thampi:2014b,Gao:2014}.

Ordered liquid crystalline phases of active matter can be classified according to their symmetry and to the nature of the forces that the active units exert on the environment.
Active particles are often elongated objects with a head and a tail, hence intrinsically polar, such as bacteria or birds.  Such systems can order in states with ferromagnetic (polar) order, where all units are on average aligned in a fixed direction. In this case the ordered state is also a macroscopically moving state. Polar active particles can also order in states with nematic or apolar order, where the particles are aligned along the same axis, but with random head/tail orientation. In this case the ordered state has apolar or nematic symmetry: if the direction of mean order is denoted by a unit vector $\bm{n}$ then the ordered state is invariant under $\bm{n}\rightarrow -\bm{n}$ and has zero mean velocity. Some active units are intrinsically apolar, such as vibrated rods~\cite{Narayan:2007}, melanocytes~\cite{Kemkemer:2000} and some fibroblasts~\cite{Duclos:2014}, and order in states with nematic symmetry. 

In addition to the distinction based on symmetry of the ordered state, active systems can be further classified by the type of stresses or flows they impose on the surrounding medium. These can be contractile, as in actomyosin networks or in migrating cell layers, or extensile, as in suspensions of microtubule bundles or in most bacteria~\cite{Marchetti:2013}. In the context of swimming microorganisms or artificial swimmers, units that exert extensile forces on the surrounding fluid are known as pushers, while those that exert contractile forces are known as pullers. Most bacteria are pushers, while the alga Chlamydomonas is an example of a puller. The extensile or contractile nature of active stresses affects the stability of ordered states~\cite{Ezhilan:2013}.

In this paper we focus on the rich dynamics of active liquid crystals with nematic symmetry. We consider both extensile~\cite{Sanchez:2012,Zhou:2014} and contractile~\cite{Duclos:2014} systems.  Previous work has highlighted the rich dynamics that arises in active nematics from the interplay of activity, orientational order and flow~\cite{Voituriez:2005,Marenduzzo:2007,Giomi:2008,Giomi:2011,Elgeti:2011,Giomi:2012}. More recently it was suggested that topological defects play an  important role in mediating and driving turbulent-like active flows~\cite{Sanchez:2012,Giomi:2013,Thampi:2013,Thampi:2014a,Thampi:2014b,Gao:2014}. 

Topological defects  are inhomogeneous configurations of the order field that provide a distinctive signature of liquid crystalline order and have been extensively studied in  passive nematics. For passive nematics, defects  may be generated through boundary conditions, externally applied fields, or via sufficiently rapid quenches from the disordered to the ordered state~\cite{Kleman:2006,Kleman:2003,Lavrentovich:1998}. When the constraints are removed, or the system is given time to equilibrate, the defects ultimately annihilate and the system reaches a homogeneous ordered state that minimizes the free energy. The structure of the topological defects is intimately related to the broken symmetry of the ordered state and effectively provides a ``fingerprint" of such a symmetry.  When the ordered state has ferromagnetic (polar) symmetry, the lowest energy defect configurations are the charge $+1$ vortices and asters (monopoles), while in states with nematic  symmetry charge $\pm 1/2$ defects, known as disclinations, are possible~\cite{DeGennes:1993} and have the lowest energy. The structure of the topological defects therefore provides an important tool for classifying the broken symmetry of liquid crystalline states. 

In active liquid crystals, in contrast to passive ones,  defect configurations can occur spontaneously in the bulk and be continuously regenerated by the local energy input, as demonstrated in experiments~\cite{Kemkemer:2000,Sanchez:2012,Zhou:2014,Duclos:2014}. While the aster and vortex defects that occur in polar active systems~\cite{Nedelec:1997,Surrey:2001} have been studied for some time \cite{Kruse:2004,Kruse:2006,Voituriez:2006,Elgeti:2011}, the properties of defects in active nematics have only recently become the focus of experimental and theoretical attention. Disclinations have been identified in monolayers of vibrated granular rods~\cite{Narayan:2007}, in active nematic gels assembled {\em in vitro} from microtubules and kinesins~\cite{Sanchez:2012},  in dense cell monolayers~\cite{Kemkemer:2000,Duclos:2014}, and in living liquid crystals obtained by injecting bacteria in chromonic liquid crystals~\cite{Zhou:2014}.  In bulk suspensions of microtubule bundles the defects drive spontaneous flows~\cite{Sanchez:2012}. When confined at an oil/water interface, furthermore, the same suspensions form a two-dimensional active nematic film, with self-sustained flows resembling cytoplasmic streaming and the continuous creation and annihilation of defect pairs \cite{Sanchez:2012}. 

Recent work by us ~\cite{Giomi:2013} and others~\cite{Thampi:2013,Pismen:2013,Thampi:2014a,Thampi:2014b,Gao:2014} has begun to systematically examine the effect of activity on the dynamics of disclinations in  a nematic liquid crystalline film.  We demonstrated that $+1/2$ disclinations in active liquid crystals behave like self-propelled particles with an active speed proportional to activity. The direction of active motion is controlled by the extensile or contractile nature of the active stresses. For certain relative orientations of pairs of defects of opposite strength this self propulsion can overcome the equilibrium repulsive interaction among pairs of opposite-sign defects, allowing for dynamical states with an average sustained concentration of defect-antidefect pairs. In related work, Thampi \emph{et al}.~\cite{Thampi:2013} suggested that the mean distance between defects in these turbulent states may be strongly correlated with the correlation length of fluctuations in the flow velocity, and only weakly dependent on activity. In the rest of this paper we first review the hydrodynamic description of  active nematics and the instabilities of the homogeneous ordered state. In Section \ref{sec:proliferation} we present the results of  a systematic numerical study of the various flow regimes induced by activity. Each regime is characterized in terms of flow patterns and defect proliferation. The chaotic regime exhibits a steady number of  defects that persist in time.   This is made possible by active flows that drive directed motion of the comet-like $+1/2$ defects, generating,  for certain relative orientations of two opposite sign defects, an effective repulsive interaction between  the pair. The mechanisms for this active defect dynamics are analyzed in Sections \ref{sec:isolated} and \ref{sec:annihilation}, where we discuss  individual defect dynamics  and pair annihilation in active nematics, respectively. We conclude with a brief discussion  highlighting open questions.

\section{\label{sec:nematodynamics}Active nematodynamics}

\subsection{\label{sec:equations}Governing Equations}

We consider a uniaxial active nematic liquid crystal in two dimensions. The two-dimensional limit is appropriate to describe the experiments by Sanchez \emph{et al}.~\cite{Sanchez:2012}, where the microtubule bundles confined to a water-oil interface form an effectively two-dimensional dense nematic  suspension, but also of considerable interest in its own right. The hydrodynamic equations of  active nematic liquid crystals have been derived by coarse-graining a semi-microscopic model of cytoskeletal filaments crosslinked by clusters of motor proteins~\cite{Liverpool:2008}. They can also simply be obtained from the hydrodynamic equations of passive systems by the addition of nonequilibrium stresses and currents due to activity \cite{Kruse:2004,Kruse:2005,Marenduzzo:2007,Joanny:2007,Giomi:2011,Giomi:2012,Marchetti:2013}. We consider here an incompressible suspension where the total density $\rho$ of active bundles and solvent is constant.  The equations are formulated in terms of the concentration $c$ of active units, the flow velocity $\bm{v}$ of the suspension and the nematic tensor order parameter $Q_{ij}=S\left(n_in_j-\delta_{ij}/2\right)$, with $\bm{n}$ the director field.  The alignment tensor $Q_{ij}$ is traceless and symmetric, and hence has only two independent components in two dimensions. The constraint of constant density $\rho$ requires  $\nabla\cdot\bm{v}=0$. The hydrodynamic equations are given by~\cite{Giomi:2011}
\begin{subequations}\label{eq:hydrodynamics}
\begin{gather}
\label{eq:c}
\frac{Dc}{Dt}=\partial_i\left[D_{ij}\partial_jc+\alpha_{1}c^{2}\partial_jQ_{ij}\right]\;,\\[7pt]
\label{eq:v}
\rho \frac{Dv_i}{Dt}=\eta{\color{black} \nabla^2} v_i-\partial_ip+\partial_j\sigma_{ij}\;,\\[5pt]
\frac{D{Q}_{ij}}{Dt}=\lambda Su_{ij}+Q_{ik}\omega_{kj}-\omega_{ik}Q_{kj}+\frac{1}{\gamma}\,H_{ij}\;,
\label{eq:Q}
\end{gather}
\end{subequations}
where $D/Dt=\partial_{t}+\bm{v}\cdot\nabla$ indicates the material derivative, $D_{ij}=D_0\delta_{ij}+D_1Q_{ij}$ is the anisotropic diffusion tensor, $\eta$ the viscosity, $p$ the pressure, $\lambda$ the nematic alignment parameter, and $\gamma$ the rotational viscosity. Here $u_{ij}=(\partial_iv_j+\partial_jv_i)/2$ and  $\omega_{ij}=(\partial_iv_j-\partial_jv_i)/2$ are the strain rate and vorticity tensor, respectively, representing the symmetric and antisymmetric parts of the velocity gradient. The molecular field $H_{ij}=-\delta F_{\rm LdG}/\delta Q_{ij}$ embodies the relaxational dynamics of the nematic obtained from the variation of the two-dimensional Landau-de Gennes free energy $F_{\rm LdG}=\int dA\,f_{\rm LdG}$ \cite{DeGennes:1993}, with
\begin{equation}
\label{eq:free_energy}
f_{\rm LdG}=\tfrac{1}{2}A\tr(\bm{Q})^{2}+\tfrac{1}{4}C(\tr\bm{Q}^{2})^{2}+\tfrac{1}{2}K|\nabla\bm{Q}|^{2}\;,
\end{equation}
where $K$ is an elastic constant with dimensions of energy. For simplicity we restrict ourselves here to the  one-elastic constant approximation to the Frank free energy: i.e. equal bend and splay moduli. The coefficients $A$ and $C$ determine the location of the continuous transition from a homogeneous isotropic state with $S=0$ to a homogeneous nematic state with a finite value of $S$ given by $S=\sqrt{-2A/C}$ (where we have used $\tr\bm{Q}^{2}=S^{2}/2$). We are interested here in a system where the transition is driven by the concentration of nematogens, as is the case for a fluid of hard rods of length $\ell$ which exhibits an isotropic-nematic (IN) transition at a concentration $c^\star=3\pi/2\ell^2$ in two dimensions. Noting that $A$ and $C$ have dimensions of energy density (in two dimensions), we choose $A=K\left(c^{\star}-c\right)/2$ and $C=Kc$ \cite{Giomi:2012}. This gives $S=S_0=\sqrt{1-c^\star/c_0}$ in a homogeneous state of density $c_0$, so that $S_0= 0$ for $c_0 < c^\star$ and $S_0\approx 1$ for $c_0\gg c^\star.$ The ratio $\sqrt{K/|A|}$ defines a length scale that corresponds to the equilibrium correlation length of order parameter fluctuations and diverges at the continuous IN transition~\cite{Kleman:2003}. Here we restrict ourselves to mean values of concentration well above $c^\star$, where this equilibrium correlation length is microscopic and is of the order of the size $\ell$ of the nematogens, which will be used as our unit of length in the numerical simulation. Finally, the stress tensor $\sigma_{ij}=\sigma^{\rm r}_{ij}+\sigma^{\rm a}_{ij}$ is the sum of the elastic stress due to nematic elasticity, 
\begin{equation}
\label{eq:sigmar}
\sigma^{\rm r}_{ij}=-\lambda S H_{ij}+Q_{ik}H_{kj}-H_{ik}Q_{kj}\;,
\end{equation}
where for simplicity we have neglected the Eriksen stress, and an active contribution, given by~\cite{Marchetti:2013}
\begin{equation}
\label{eq:sigmaa}
\sigma^{\rm a}_{ij}=\alpha_{2}c^{2}Q_{ij}\;,
\end{equation}
which describes stresses exerted by the active particles.  The sign of $\alpha_{2}$ depends on whether the active particles generate contractile or extensile stresses, with $\alpha_{2}>0$ for the contractile case and $\alpha_{2}<0$ for extensile systems. Activity yields also a curvature-induced current given by the last term on the right hand side of Eq.~\eqref{eq:c}, $\bm{j}^{\rm a}=-\alpha_{1}c^{2}\nabla\cdot\bm{Q}$, that drives active units from regions populated by fast moving particles to regions of slow moving particles. The $c^{2}$ dependence of the active stress and current is appropriate for systems where activity arises from pair interactions among the filaments via crosslinking motor proteins.  The active parameters $\alpha_1$ and $\alpha_2$ will be treated here as purely phenomenological. In microscopic models they are found to depend on the concentration of active crosslinkers and on the consumption rate of adenosine triphosphate (ATP)~\cite{Liverpool:2003,Liverpool:2005}.
 
\subsection{\label{sec:linearstability}Linear Stability}

The hydrodynamic equations of an active nematic have two homogeneous stationary solutions, with $c=c_0$ and $\bm{v}=\bm{0}$. For $c_0<c^\star$ the homogeneous state is disordered with $S_0=0$. For $c_0>c^\star$ the homogeneous solution is an ordered nematic state with $S_0=\sqrt{1-c^\star/c_0}$. The linear stability of the ordered state has been studied in detail for the case of a $L\times L$ periodic domain \cite{Giomi:2011,Giomi:2012}. It is found that above a critical activity the homogenous state is unstable to a laminar flowing state. This instability corresponds to the  spontaneous flow instability well-studied in a channel geometry \cite{Voituriez:2005,Giomi:2008,Edwards:2009}. The critical activity value associated with the instability of the homogeneous state is given by~\cite{Giomi:2012}
\begin{equation}\label{eq:alphac}
\alpha_{2}^{\pm}=\pm \frac{4\pi^2 K [2\eta+\gamma S_0^{2}\left(1\mp\lambda\right)^{2}]}{\gamma c_{0}^{2}L^{2}S_0\left(1\mp\lambda\right)}\;.
\end{equation}

\begin{figure}
\includegraphics[width=0.8\columnwidth]{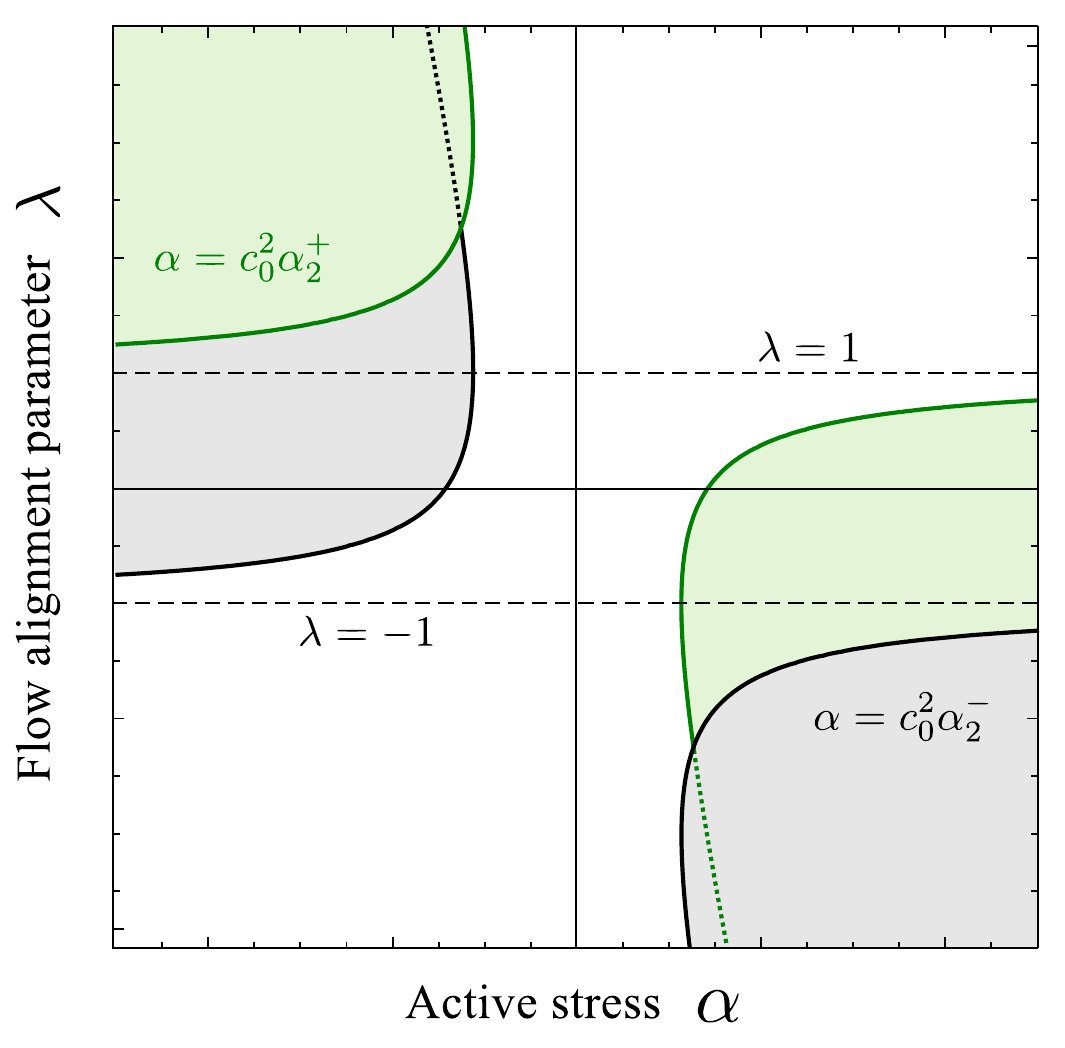}
\caption{\label{fig:stability}Schematic representation of the region where an active nematic is linearly unstable to splay (green) and to bend (gray) fluctuations in the plane of the alignment parameter $\lambda$ and the activity $\alpha=c_0^2\alpha_2$. The unstable regions are bounded by the critical activity given in Eq.~\eqref{eq:alphac}. Flow tumbling extensile nematic with $|\lambda|<1$ are unstable to bend when active stresses are extensile ($\alpha<0$) and to splay when active stresses are contractile ($\alpha>0$).  Conversely, strongly flow aligning ($|\lambda|\gg1$) are unstable to splay when active stresses are extensile ($\alpha<0$) and to bend when active stresses are contractile ($\alpha>0$). }
\end{figure}

\noindent For $|\alpha_{2}|>|\alpha_{2}^{+}|$ the system is subject to spontaneous splay deformations characterized by the instability of the first transverse mode. For $|\alpha_{2}|>|\alpha_{2}^{-}|$, on the other hand, the instability is determined by the first longitudinal mode corresponding to bending deformations. Note that the sign of $\alpha_{2}^{\pm}$ depends both on the sign of the fraction and the sign of the term $1\mp\lambda$ in the denominator. Thus flow-aligning nematics ($|\lambda|>1$) are unstable to splay under the effect of an extensile active stress ($\alpha_{2}<\alpha_{2}^{+}<0$) and to bending under the effect of a contractile active stress ($\alpha_{2}>\alpha_{2}^{-}>0$). Vice versa, flow-tumbling nematics ($-1<\lambda<1$) are unstable to bending under the effect of an extensile active stress ($\alpha_{2}<\alpha_{2}^{-}<0$) and to splay under the effect of a contractile active stress ($\alpha_{2}>\alpha_{2}^{+}>0$) (see Fig.\ref{fig:stability}). In this paper we focus exclusively on flow-tumbling systems and discuss both the cases of extensile and contractile stresses.

\subsection{\label{sec:numerics}Dimensionless Units and Numerical Methods}

To render Eqs.~\eqref{eq:hydrodynamics} dimensionless, we scale distances by the length of the active nematogens $\ell$, set by the critical concentration $c^{\star}$, stresses by the elastic stress of the nematic phase $\sigma=K/\ell^{2}$ and time by $\tau_p=\eta\ell^{2}/K$ representing the ratio between viscous and elastic stress. In these dimensionless units we take $\alpha_{1}=|\alpha_{2}|/2$ and we introduce
\begin{equation}\label{eq:active_stress_scale}
\alpha = \alpha_{2}c_{0}^{2}\;,	
\end{equation}
as the fundamental measure of active stress. This will serve as the control parameter throughout this work. 

The numerical calculations presented in Sec. \ref{sec:annihilation} and \ref{sec:proliferation} are performed via finite differences on a square grid of $256^{2}$ points. The time integration was performed via a fourth order Runge-Kutta method with time step $\Delta t = 10^{-3}$. Except where mentioned otherwise, the numerical calculations described in this section use the parameter values   $ D_{0} = D_{1} = 1$, $\lambda = 0.1$, $c_{0} = 2c^\star$ (corresponding to $S_0=0.707$) and $L = 20$.

\section{\label{sec:isolated}Dynamics of an isolated disclination}

Topological defects are spatially inhomogeneous configurations of the director field that cannot be transformed continuously into a uniform state.  In equilibrium defects can occur upon quench from the disordered into the ordered phase or upon application of external electric or magnetic fields. Geometry or suitable boundary conditions can also be used to generate and maintain defects in the system. For instance in a spherical nematic droplet, with boundary condition such that the director is normal to the droplet surface, the equilibrium configuration consists of a radial director field with a point defect at the center of the droplet. In the absence of such constraints or external fields and given enough time to equilibrate defects of opposite sign always annihilate and the system settles into a uniform equilibrium state \cite{Chuang:1993,Bray:2002}. 

In two dimensions defects are point-like. The strength of a disclination depends on how much the director field rotates around the defect core in one  loop. In two dimensions this can be expressed in terms of a single scalar field $\theta$ representing the angle formed by the director $\bm{n}=(\cos\theta,\sin\theta)$ with the horizontal axis of a Cartesian frame. This gives
\begin{equation}\label{eq:contour_integral}
\oint d\theta = 2\pi k\;,	
\end{equation}
where the integral is calculated along an arbitrary contour enclosing the defect. The integer $k$ is called strength of the defect and is analogous to the winding number of vortex defects in polar systems. In two-dimensional uniaxial nematics the lowest energy defect configurations consists of half-strength disclinations with $k=\pm 1/2$. In the presence of an isolated defect located at the origin, a solution $\theta=\theta_{\rm d}$ that minimizes the energy \eqref{eq:free_energy} and satisfies the constraint \eqref{eq:contour_integral} is given by
\begin{equation}\label{eq:defective_theta}
\theta_{\rm d}=k\phi\;, 
\end{equation}
 with $\phi$ the usual polar angle. The corresponding energy is given by
\begin{equation}\label{eq:energy}
F = \pi K k^{2} \log(R/a)+\epsilon_{\rm c}\;,
\end{equation}
where $R$ is the size of the system and $a$ is the core radius, defined as the radius of the region in the immediate proximity of the defect where the order parameter drops from its equilibrium value to zero. The quantity $\epsilon_{\rm c}$ is the  contribution to the total energy due to the isotropic defect core. 

Point defects in nematic liquid crystals have several particle-like features. Like charged particles, opposite-sign defects attract and same-sign defects repel (\cite{DeGennes:1993} and Sec. \ref{sec:annihilation}). It is well established that the dynamics of an isolated disclination that evolves according to Eqs. \eqref{eq:hydrodynamics} can be cast in the form of an overdamped  equation of motion, given by
\begin{equation}\label{eq:newton}
\zeta \left(\frac{d\bm{r}}{dt}-\bm{v}\right) = \bm{F}\;,
\end{equation}
where $\bm{r}$ is the defect position, $\bm{F}$ is the net force acting on the defect, due to interaction with other defects or externally imposed perturbations, and $\bm{v}$ the local flow velocity at the position of the defect, which includes both external and self-generated flows. Finally, $\zeta$ is an effective drag coefficient proportional to the rotational viscosity $\gamma$ in Eq. (\ref{eq:hydrodynamics}c) and possibly space-dependent.

In the absence of fluid flow an isolated disclination moves only in response to an externally imposed distortion and relaxes to the minimal energy texture $\theta_{\rm d}$ given in Eq. \eqref{eq:defective_theta}. Following Denniston \cite{Denniston:1996}, one can then set $\theta=\theta_{\rm d}+\theta_{\rm ext}$, where $\theta_{\rm ext}$ expresses the departure from the optimal defective configuration $\theta_{\rm d}$, and calculate the energy variation with respect to a small virtual displacement of the defect core. For small deformations, this gives $\bm{F}=-2\pi k K \nabla_{\perp}\theta_{\rm ext}$, with $\nabla_{\perp}=(-\partial_{y},\partial_{x})$, and 
\begin{equation}\label{eq:effective_friction}
\zeta = \pi \gamma k^{2} \int_{{\rm Er}}^{\infty} dx\,K_{1}^{2}(x)I_{1}(2x) \approx 226\,\pi \gamma k^{2}\;,
\end{equation}
where $K_{1}$ and $I_{1}$ are Bessel functions and ${\rm Er}=\gamma a |\dot{\bm{r}}| / K$ is the Ericksen number at the length scale of the defect core. The second equality in Eq. \eqref{eq:effective_friction} implies $a\approx 0$; for finite core radius it introduces a dependence of the effective friction $\zeta$ on the defect velocity. We refer the reader to Refs. \cite{Denniston:1996,Pleiner:1988,Ryskin:1991} for details.  

The hydrodynamic coupling between the local orientation of the director  and the flow gives rise to a self-generated flow, known as \emph{backflow}, that in turn advects the defect core. 
When the dynamics of the flow is much faster than the orientational dynamics of the director, and in the absence of external forces, one can neglect  $\bm{F}$ in Eq.~\eqref{eq:newton} and calculate the flow velocity $\bm{v}$ from the Stokes equation,
\begin{equation}\label{eq:stokes} 
\eta\nabla^{2}\bm{v}-\nabla p + \bm{f} = \bm{0}\;,\qquad 
\nabla\cdot\bm{v} = 0\;,
\end{equation}
where $\bm{f}=\bm{f}^{\rm a}+\bm{f}^{\rm e}=\nabla\cdot(\bm{\sigma}^{\rm a}+\bm{\sigma}^{\rm r})$ is the force arising from the active and elastic stresses acting in the system. In the case of  isolated defects this scenario is generally realistic for $\eta/\gamma \ll 1$ and even in a system containing multiple defects this purely advective dynamics continues to hold as long as the defects are sufficiently far apart (see Sec. \ref{sec:annihilation}). The general case in which both $\bm{v}$ and $\bm{F}$ are non-zero was discussed by Kats {\em et al}. \cite{Kats:2002}. 

\begin{figure}[t]
\centering
\includegraphics[width=\columnwidth]{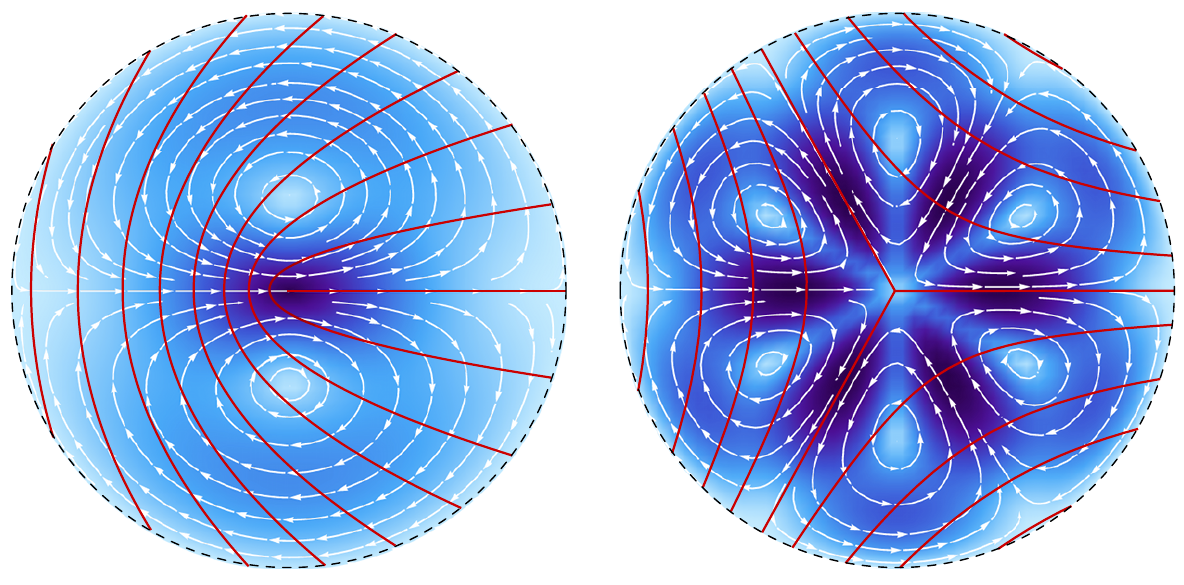}
\caption{\label{fig:defect_flows}Example of a $+1/2$ (left) and $-1/2$ (right) disclination. The solid red lines are tangent to the director field $\bm{n}=(\cos\theta_{\rm d},\sin\theta_{d})$ with $\theta_{d}=k\phi$ and $k=\pm 1/2$. The background shows the active backflow associated with the disclinations and obtained by solving the Stokes equation \eqref{eq:stokes} with no-slip boundary conditions on a circle (dashed black line). The intensity of the background color is proportional to the magnitude of the flow velocity. The white streamlines are given by Eq. \eqref{eq:active_backflow}.}	
\end{figure}

In the remainder of this section we consider the regime in which $\eta/\gamma \ll 1$ and calculate the backflow due to the stresses arising in the presence of  isolated $k=\pm 1/2$ disclinations. Let us then consider a $\pm 1/2$ disclination located at the origin of a circular domain of size $R$. The domain might represent either the entire system or, more realistically, the defect-free portion of the system surrounding a given central defect. We will refer to this as the \emph{range} of a defect. Because of the linearity of the Stokes equation, the solution of Eq. \eqref{eq:stokes} can be  written as $\bm{v}=\bm{v}^{0}+\bm{v}^{\rm a}+\bm{v}^{\rm e}$, where $\bm{v}^{0}$ is the solution of the homogeneous Stokes equation, while $\bm{v}^{\rm a}$ and $\bm{v}^{\rm e}$ are the flows produced by the active and elastic force, respectively. The solution can be expressed as the convolution of the two-dimensional Oseen tensor with the force per unit area,
\begin{equation}\label{eq:general_solution}
v_{i}(\bm{r}) = \int dA'\,G_{ij}(\bm{r}-\bm{r}')f_{j}(\bm{r}')\;,	
\end{equation}	
where $G_{ij}$ is the two-dimensional Oseen tensor \cite{DiLeonardo:2008}, given by
\begin{equation}\label{eq:oseen}
G_{ij} (\bm{r}) = \frac{1}{4\pi\eta}\,\left[\left(\log\frac{\mathcal{L}}{r}-1\right)\delta_{ij}+\frac{r_{i}r_{j}}{r^{2}}\right]\;,	
\end{equation}	
with $\mathcal{L}$ a length scale adjusted  to obtain the desired behavior at the boundary. Taking $\bm{n}=(\cos k\phi,\sin k\phi)$, with $k=\pm 1/2$, and assuming uniform concentration and nematic order parameter outside the defect core, the body force due to activity can be calculated straightforwardly as
\[
\bm{f}^{\rm a}
= \nabla\cdot\bm{\sigma}^{\rm a}	
= \frac{\alpha}{2r}
\left\{
\begin{array}{lll}
\bm{\hat{x}} & & k=+1/2\;, \\[7pt]
-\cos 2\phi\,\bm{\hat{x}}+\sin 2\phi\,\bm{\hat{y}} & & k=-1/2\;,	
\end{array}
\right.
\]
where, for simplicity, we have assumed $S_{0}=1$ outside the core. Using this in Eq. \eqref{eq:general_solution}, and using \eqref{eq:oseen}, yields, after some algebraic manipulation,
\begin{widetext}
\begin{subequations}\label{eq:active_backflow}
\begin{gather}
\bm{v}_{+}^{\rm a}(r,\phi) = \frac{\alpha}{12\eta}\left\{[3(R-r)+r\cos 2\phi]\,\bm{\hat{x}}+r\sin 2\phi\,\bm{\hat{y}}\right\}\;,\\[10pt]
\bm{v}_{-}^{\rm a}(r,\phi) 
= \frac{\alpha r}{12 \eta R}\left\{
\left[\left(\frac{3}{4}\,r-R\right)\cos 2\phi-\frac{R}{5}\,\cos 4\phi\right]\bm{\hat{x}}+
\left[\left(\frac{3}{4}\,r-R\right)\sin 2\phi+\frac{R}{5}\,\sin 4\phi\right]\bm{\hat{y}} 
\right\}\,.	
\end{gather}
\end{subequations}
\end{widetext}
A plot of these flow fields is shown in Fig. \ref{fig:defect_flows}. Setting $\bm{v}=\bm{v}_{+}^{\rm a}(0,\phi)$ in Eq. \eqref{eq:newton} we find that active $+1/2$ disclinations self-propel at constant speed along their symmetry axis ($\bm{\hat{x}}$ in this setting), and their equation of motion can be written as
\begin{equation}\label{eq:spp}
\zeta \frac{d\bm{r}_{+}}{dt} = v_{0}\bm{\hat{x}}\;,
\end{equation} 
where $v_{0}=\alpha R/(4\eta)$. On the other hand, $\bm{v}_{-}^{\rm a}(0,\phi)=\bm{0}$, and so $-1/2$ disclinations are not propelled by the active backflow, but rather move solely under the effect of the elastic force produced by other defects. It is crucial to notice that the self-propulsion speed $v_{0}$ scales linearly with the active stress $\alpha$ and so disclinations in contractile ($\alpha>0$) and extensile ($\alpha<0$) active nematic suspensions self-propel in opposite directions.

Equations \eqref{eq:active_backflow} can be complemented with various kinds of boundary conditions, which, however, have no effect on the qualitative features of the solution and, more importantly, on the dynamics of the defects. To illustrate this point we consider a slippery interface, such that $v_{r}(R,\phi)=0$ and $\sigma^{\rm a}_{r\phi}=-\xi v_{\phi}(R,\phi)$, where $\xi$  is the coefficient associated with the frictional force exerted by the interface on the fluid. If the domain of Eq. \eqref{eq:stokes} is interpreted as a container, then $\xi$ is the actual frictional coefficient of the container wall. On the other hand, if the domain is interpreted as the \emph{range} of a defect, then $\xi\sim\eta h$, where $h$ is the thickness of the boundary layer between the ranges of neighboring defects. No-slip boundary conditions can be recovered in the limit $\xi\rightarrow\infty$.
 
A solution $\bm{v}^{0}$ of the homogeneous Stokes equation enforcing the boundary conditions for the active backflow induced by the $+1/2$ disclination can be found from the following biharmonic stream function
\begin{equation}\label{eq:positive_stream}
\psi_{+} = (a_{1}r+b_{1}r^{3})\sin\phi\;,	
\end{equation}
with $a_{1}$ and $b_{1}$ constants.	The corresponding velocity field $\bm{v}_{+}^{0}=(\partial_{y}\psi,-\partial_{x}\psi)$ is given by
\begin{equation}
\bm{v}_{+}^{0}(r,\phi) = [a_{1}+b_{1}r^{2}(2-\cos 2\phi)]\,\bm{\hat{x}}-b_{1}r^{2}\sin 2\phi\,\bm{\hat{y}}\;. 
\end{equation}
Then, setting $\bm{\hat{r}}\cdot\bm{v}_{+}(R,\phi)=0$ and $\bm{\hat{\phi}}\cdot\bm{v}_{+}(R,\phi)=-\sigma^{\rm a+}_{r\phi}(R,\phi)/\xi=\alpha/(2\xi)\sin\phi$, with $\bm{v}_{+}=\bm{v}_{+}^{0}+\bm{v}_{+}^{\rm a}$, and solving for $a_{1}$ and $b_{1}$ yields
\[
a_{1} = -\frac{\alpha R}{6\eta}+\frac{\alpha}{4\xi}\;,\qquad
b_{1} = \frac{\alpha}{12 \eta R}-\frac{\alpha}{4\xi R^{2}}\;.
\]
The associated self-propulsion speed $v_{0}$ in Eq. \eqref{eq:spp} then becomes 
\begin{equation}
v_{0} = \alpha\left(\frac{R}{12\eta}+\frac{1}{4\xi}\right)\;,
\end{equation}
or $v_{0}=\alpha R/(12\eta)$ in the no-slip limit. Thus, as anticipated, incorporating the effect of the boundary only changes the speed of the defects without altering their  dynamics. Similarly, in the case of a $-1/2$ disclination, we can consider the biharmonic stream function  
\begin{equation}\label{eq:negative_stream}
\psi_{-} = (a_{3}r^{3}+b_{3}r^{5})\sin 3\phi\;, 	
\end{equation}
whose associated velocity field is given by
\begin{multline}
\bm{v}_{-}^{0}(r,\phi) 
= \left[r^{2} (3a_{3}+4 b_{3} r^{2})\cos 2\phi-b_{3}r^{4}\cos 4\phi\right]\,\bm{\hat{x}}\\
- \left[r^{2} (3a_{3}+4 b_{3} r^{2})\sin 2\phi+b_{3}r^{4}\sin 4\phi\right]\,\bm{\hat{y}}\;.
\end{multline}
Setting $\bm{\hat{r}}\cdot\bm{v}_{-}(R,\phi)=0$ and $\bm{\hat{\phi}}\cdot\bm{v}_{-}(R,\phi)=-\sigma^{\rm a-}_{r\phi}(R,\phi)/\xi=\alpha/(2\xi)\sin 3\phi$ and solving for $a_{3}$ and $b_{3}$ yields
\[
a_{3} = \frac{7\alpha}{240\eta R}+\frac{\alpha}{4\xi R^{2}}\;,\qquad
b_{3} =-\frac{\alpha}{60\eta R^{3}}-\frac{\alpha}{4\xi R^{4}}\;.
\]
Clearly this does not change the symmetry of the active backflow, and thus negative-charge disclinations are stationary. 

In summary, half-strength disclinations in active nematic liquid crystals can be described as  self-propelled particles with overdamped dynamics governed by an equation of the form \eqref{eq:newton}. In the absence of external forces, or forces due to interactions with other defects, the disclination core is advected at constant speed by a self-generated backflow, provided the dynamics of the flow is faster then the relaxational dynamics of the director (i.e. $\eta/\gamma\ll 1$). As a result, positive disclinations travel along their symmetry axis at speed $v_{0} \sim \alpha R/\eta$, while negative disclinations are stationary and move only as a consequence of their interaction with other disclinations. The direction of motion is controlled by the sign of the activity $\alpha$, which in turns depends on whether the system is contractile ($\alpha>0$) or extensile ($\alpha<0$). Thus the comet-like $+1/2$ disclination travels in the direction of its ``tail'' in contractile systems and in the direction of its ``head'' in extensile systems. We note that active curvature currents in the concentration equation controlled by the parameter $\alpha_{1}$ have a similar effect, as  noted by Narayan {\em et al}. in a system of vibrated rods \cite{Narayan:2007} and recently investigated by Shi and Ma through extensive numerical simulations \cite{Shi:2013}. Such curvature driven currents control the dynamics in systems with no momentum conservation, but are very small in the regime discussed here.

Special attention should be devoted to the fact that the self-propulsion speed $v_{0}$ depends linearly on the range $R$ of a defect, this being defined as the defect-free portion of the system surrounding a defect (and possibly coinciding with the entire system).  Although two-dimensional hydrodynamics is known to be plagued with  anomalies, such as the Stokes paradox \cite{Lamb:1945}, this behavior does not result solely from the two-dimensionality of the problem.  Point defects in three dimensions would also yield a similar behavior. To see this we note that the deviatoric part of the Oseen tensor scales like $r^{2-D}$, with $D$ the space dimension. On the other hand, the active force $\bm{f}^{\rm a}=\alpha\nabla\cdot\bm{Q}$ always scales like $r^{-1}$. The backflow velocity in $D$ dimensions thus scales like $v^{\rm a}\sim\int_{0}^{R}d^{D}r\,r^{2-D}r^{-1}=R$, regardless of the space dimension. In three dimensions disclinations are, however,  line defects. In this case, denoting by $\xi_{d}$ the persistence length of the disclinations, i.e., the length scale over which these line defects can be treated as straight lines, the length $R$ controlling the flow velocity induced by a $+1/2$ defect would scale as $R\sim \xi_{d}\log(L/a)$, with $L$ the system size and $a$ the core radius. These results could also be obtained on the basis of dimensional analysis by noting that $\bm{v}^{\rm a}$ is always proportional to $\alpha/\eta$. Since $\alpha$ has dimensions of stress and $\eta$ has dimensions of stress over time, the resulting velocity must also be proportional to a length scale. This length scale was first noted experimentally by Sanchez {\em et al.} \cite{Sanchez:2012} and investigated numerically by Thampi {\em et al}. \cite{Thampi:2013,Thampi:2014b}. It's nature, however, remains elusive (see Sec. \ref{sec:proliferation} for further discussion on this matter).

The growth of the flow field generated by a defect at large distances can also be cut off by a frictional force $\bm{f}_{\rm s}=-\zeta_{\rm s}\bm{v}$ as may arise from the fact that the nematic film is confined at an oil/water interface~\cite{Lubensky:1996}. Such a frictional interaction with the subphase removes energy from the flow at the length scale $\ell_{\rm s}=\sqrt{\eta/\zeta_{\rm s}}$, thus controlling the decay of the velocity field. Finally, the limit where the friction dominates viscous forces corresponding to a no-slip Hele-Shaw geometry has been discussed in detail by Pismen~\cite{Pismen:2013}. In this case the flow generated by a single disclination is found to decay as $\sim r^{-3}$ at large distances from the defect. 

\section{\label{sec:annihilation}Annihilation dynamics of defect pairs}

\begin{figure}
\includegraphics[width=1\columnwidth]{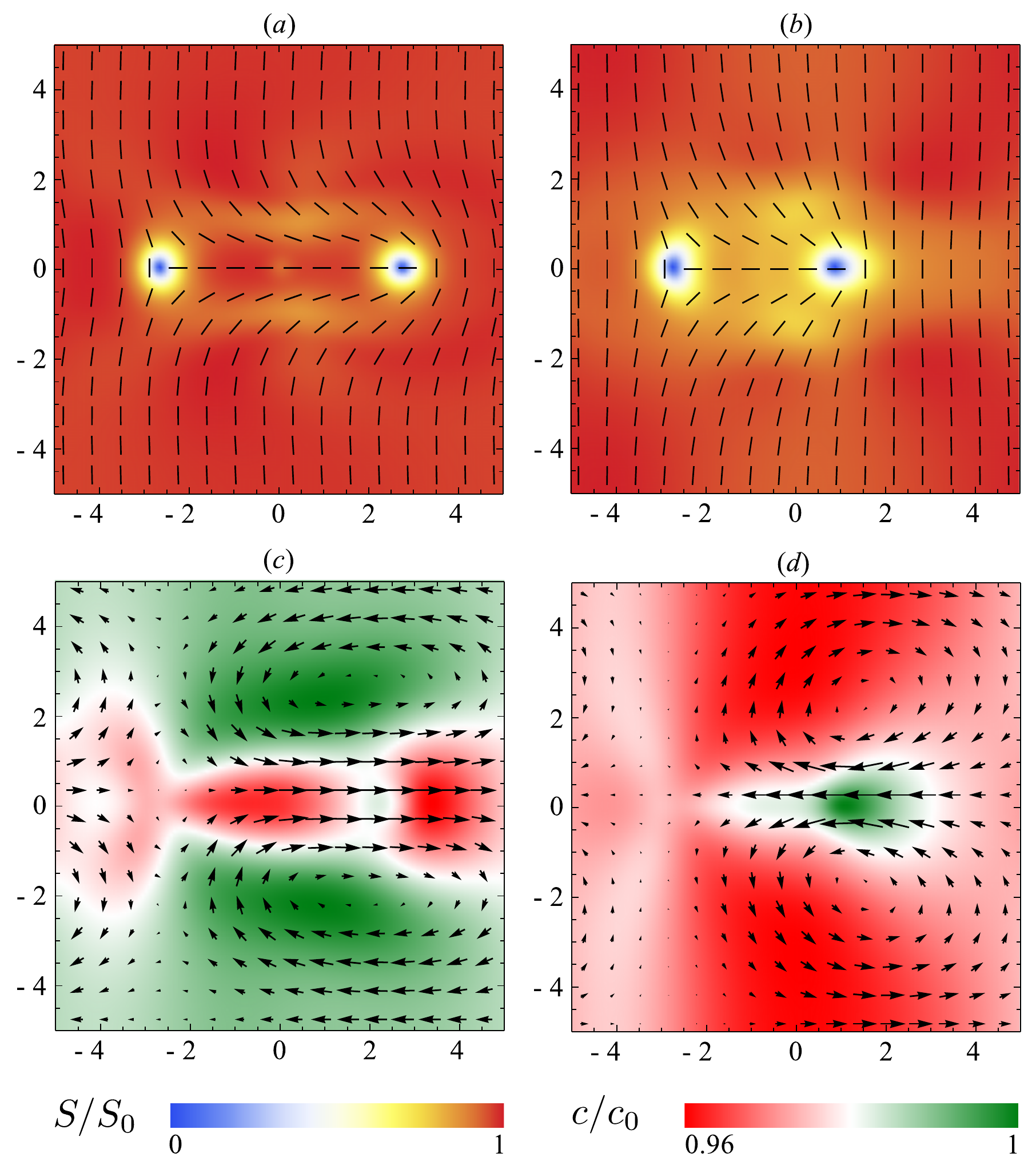}
\caption{\label{fig:annihilation}Snapshots of a disclination pair shortly after the beginning of relaxation. (Top) Director field (black lines) superimposed on a heat map of the nematic order parameter and (bottom) flow field (arrows) superimposed on a heat map of the concentration for an extensile system with $\alpha=-0.8$ (a),(c) and a contractile system with $\alpha=0.8$ (b),(d). In the top images the color denotes the magnitude of the nematic order parameter $S$ relative to its equilibrium value $S_{0}=\sqrt{1-c^{\star}/c_{0}}=1/\sqrt{2}$. In the bottom images the color denotes the magnitude of the concentration $c$ relative to the average value $c_{0}=2c^\star$. Depending on the sign of $\alpha$, the backflow tends to speed up $(\alpha>0)$ or slow down ($\alpha<0$) the annihilation process by increasing or decreasing the velocity of the $+1/2$ disclination. For $\alpha$ negative and sufficiently large in magnitude, the $+1/2$ defect reverses its direction of motion (c) and escapes  
annihilation.}
\end{figure}

In this section we discuss  the annihilation of a pair of oppositely charged  disclinations. The study of the annihilation dynamics of defect-antidefect pairs is a mature topic in the liquid crystals field and has been subject to numerous investigations \cite{Denniston:1996,Pleiner:1988,Toth:2002,Bogi:2002,Svensek:2002,Sonnet:2009,Dierking:2012}. In the simplest setting \cite{DeGennes:1993}, one considers a pair of $k=\pm 1/2$ disclinations located at $\bm{r}_{\pm}=(x_{\pm},0)$ and separated by a distance $\Delta=x_{+}-x_{-}$ (Fig. \ref{fig:annihilation}). The energy of the pair is given by
\begin{equation}\label{eq:energy_pair}
E_{\rm pair} = 2\pi k^{2} K \log(\Delta/a) + 2\epsilon_{c}\;.
\end{equation}
Each defect experiences an elastic force of the form $\bm{F}_{\pm} = -(\partial E_{\rm pair}/\partial x_{\pm})\,\hat{\bm{x}}$ and thus, in the absence of backflow, Eq. \eqref{eq:newton} can be cast in the form
\begin{equation}\label{eq:pair_force}
\frac{dx_{\pm}}{dt} = \mp \frac{\kappa}{x_{+}-x_{-}}\;,	
\end{equation}
where $\kappa=2\pi k^{2} K/\zeta$. This yields
\begin{equation}\label{eq:pair_newton}
\frac{d\Delta}{dt} = -\frac{2\kappa}{\Delta}\;,	
\end{equation}
so that the distance between annihilating defects decreases as a square-root, $\Delta(t)\propto \sqrt{t_{\rm a}-t}$, with $t_{\rm a}$ the annihilation time. More precise calculations have shown that the effective friction is itself a function of the defect separation \cite{Pleiner:1988,Ryskin:1991}, $\zeta=\zeta_{0}\log (\Delta/a)$, although this does not imply  substantial changes in the overall picture. This simple model predicts that the defect and antidefect approach each other along symmetric trajectories and annihilate at $\Delta(0)/2$ in a time $t_{\rm a}=\Delta^{2}(0)/4\kappa$. The backflow produced by the balance of elastic and viscous stresses \cite{Toth:2002,Dierking:2012}, as well as the anchoring conditions at the boundary \cite{Bogi:2002}, can produce an asymmetry in the trajectories of the annihilating defects or even suppress annihilation when the defects are initially far from each other or the anchoring is sufficiently strong.

To understand how activity changes the simple annihilation dynamics described so far, we have integrated numerically Eqs. \eqref{eq:hydrodynamics} for an initial configuration of uniform concentration and zero flow velocity, with two disclinations of charge $\pm 1/2$ symmetrically located with respect to the center of the box along the $x-$axis. Fig. \ref{fig:annihilation} shows a snapshot of the order parameter and flow field shortly after the beginning of the relaxation for both a contractile and extensile system, with $\alpha=\pm 0.8$ in the units defined in Sec. \ref{sec:numerics}. 

\begin{figure}[t]
\includegraphics[width=\columnwidth]{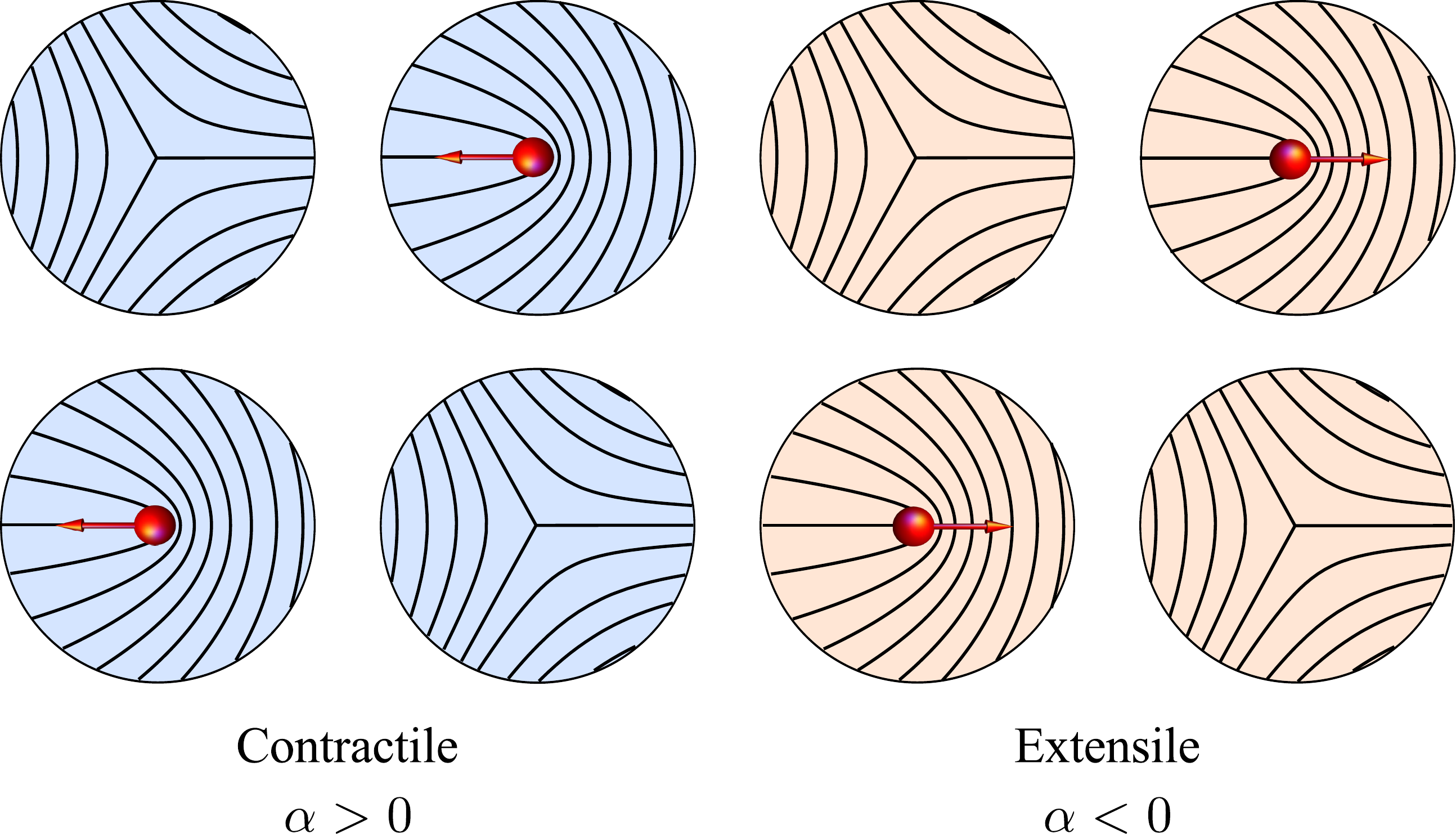}
\caption{\label{fig:pair}Schematic representation of the effective attractive/repulsive interaction promoted by the active backflow. Depending on the sign of the active stress $\alpha$, $+1/2$ disclinations self-propel in the direction of their ``tail'' (contractile)  or ``head" (extensile). Based on the mutual orientation of the defects, this can lead to an attractive or repulsive interaction.}
\end{figure}
\begin{figure}[t]
\includegraphics[width=\columnwidth]{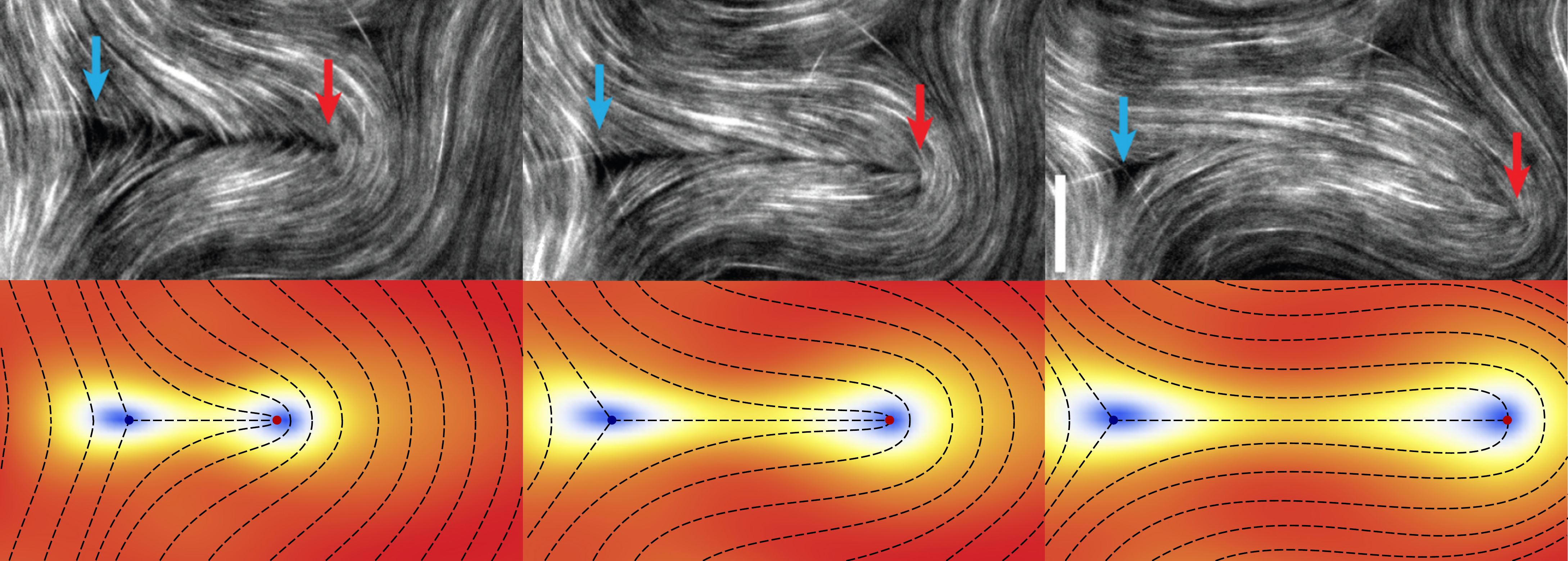}
\caption{\label{fig:pair_production}Defect pair production in an active suspension of microtubules and kinesin (top) and the same phenomenon observed in our numerical simulation of an extensile nematic fluid with $\gamma=100$ and $\alpha=-2$. The experimental pictures are reprinted with permission from T. Sanchez {\emph et al.,} Nature (London) 491, 431 (2012). Copyright 2012, Macmillan.}
\end{figure}

In contractile systems active backflow yields a net speed-up of the $+1/2$ defect towards its antidefect for the annihilation geometry shown in Fig.~\ref{fig:annihilation}b. In extensile systems, with $\alpha<0$, backflow drives the $+1/2$ defect to move towards its head, away from its $-1/2$ partner in the configuration of Fig. \ref{fig:annihilation}b, acting like an effectively {\em repulsive} interaction. If the initial positions of the defects are exchanged, the behavior is reversed. The effective attraction or repulsion between oppositively charged active defects is thus dictated by both the contractile or extensile nature of the active stresses, which determines the direction of the backflow, and the relative orientation of the defects, as summarized in Fig. \ref{fig:pair}. This effect has been observed in experiments with extensile microtubules and kinesin assemblies \cite{Sanchez:2012} and can be understood on the basis of the hydrodynamic approach embodied in Eqs. \eqref{eq:hydrodynamics}. In Fig. \ref{fig:pair_production} we have reproduced from Ref.~\cite{Sanchez:2012} a sequence of snapshots showing a pair of $\pm 1/2$ disclinations moving apart from each other together with the same behavior observed in our simulations.

\begin{figure}[t]
\includegraphics[width=\columnwidth]{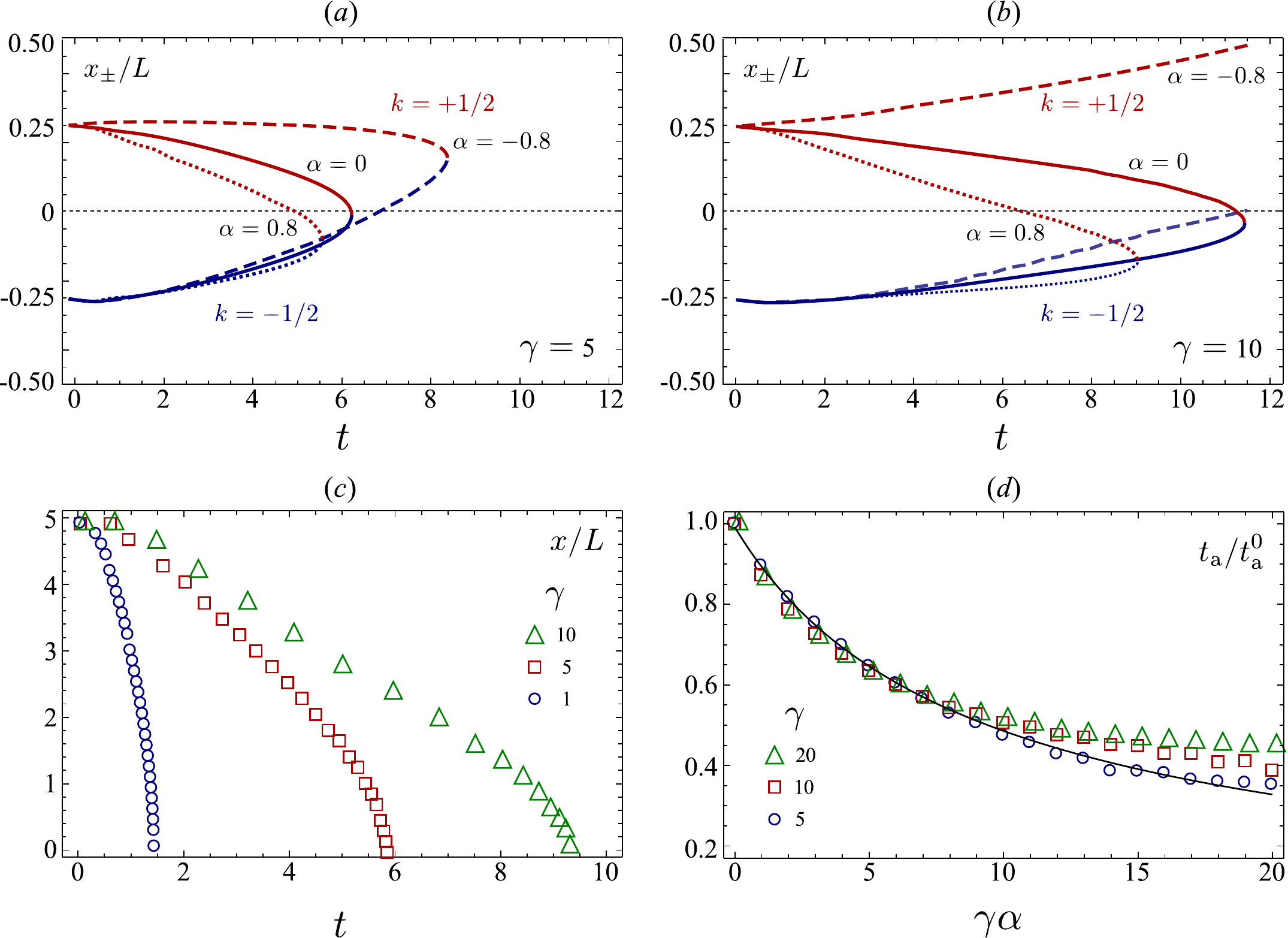}
\caption{Defect trajectories and annihilation times obtained from a numerical integration of Eqs. \eqref{eq:hydrodynamics} for various $\gamma$ and $\alpha$ values. (a) Defect trajectories for $\gamma=5$ and various $\alpha$ values (indicated in the plot). The upper (red) and lower (blue) curves correspond to the positive and negative disclination, respectively. The defects annihilate where the two curves merge. (b) The same plot for $\gamma=10$. Slowing down the relaxational dynamics of the nematic phase increases the annihilation time and for $\alpha=-0.8$ reverses the direction of motion of the $+1/2$ disclination. (c) Defect separation as a function of time for $\alpha=0.8$ and various $\gamma$ values. (d) Annihilation time normalized by the corresponding annihilation time obtained at $\alpha=0$ (i.e., $t_{\rm a}^{0}$). The line is a fit to the model described in the text. }
\label{fig:trajectories}
\end{figure}

Figs.~\ref{fig:trajectories}a and \ref{fig:trajectories}b show the trajectories of the active defects, with the red and blue line representing the $+1/2$ and $-1/2$ disclination respectively. The tracks end when the cores of the two defects merge. For small activity and small values of the rotational viscosity $\gamma$, the trajectories resemble those obtained for passive systems \cite{Toth:2002,Dierking:2012}. At large values of activity, however, the asymmetry in defect dynamics becomes more pronounced, and when the activity dominates over orientational relaxation, the $+1/2$ disclination moves independently along its symmetry axis with a speed $v_{0} \sim \alpha R/\eta$ (see Sec. \ref{sec:isolated}) whose direction is dictated by the sign of $\alpha$. This behavior is clearly visible in Fig. \ref{fig:trajectories}c, showing the defect separation $\Delta(t)$ as a function of time. For $\gamma$ sufficiently large, the trajectories are characterized by two regimes. For large separation the dynamics is dominated by the active backflow, and thus $\dot{\Delta}(t)\propto-\alpha$ and $\Delta(t) \propto -\alpha t$. Once the defects are about to annihilate, the attractive force takes over, and the defects behave as in the passive case with $\Delta(t)\propto\sqrt{t_{\rm a}-t}$.

This behavior can be understood straightforwardly from the basic concepts of active defect dynamics discussed in Sec. \ref{sec:isolated}. Each defect in the pair travels in space according to Eq. \eqref{eq:newton}, with $\bm{v}$ given by $\bm{v}(x)=\bm{v}_{+}^{\rm a}(x-x_{+})+\bm{v}_{-}^{\rm a}(x-x_{-})$ and $\bm{v}_{\pm}$ given in Eqs. \eqref{eq:active_backflow} plus a suitable homogeneous solution of the Stokes equation that enforces the periodic boundary conditions. Next, we retain only the active contribution to the backflow and replace the flow profiles by their constant values at the core of the defect, with $\bm{v}_{+}(x_{+})=v_{0}\bm{\hat{x}}\propto -\alpha$ and $\bm{v}_{-}(x_{-})=\bm{0}$. This yields the following simple equation for the pair separation
\begin{equation}
\frac{d\Delta}{dt} = v_{0} - \frac{2\kappa}{\Delta}\;.
\end{equation}
This equation explicitly captures the two regimes shown in Fig. \ref{fig:trajectories}(c) and described earlier. The solution takes the form
\begin{equation}\label{eq:distance_vs_time}
\Delta(t)=\Delta(0)+v_{0} t-\frac{2\kappa}{v_{0}}\log\left[\frac{\Delta(t)-\frac{2\kappa}{v_{0}}}{\Delta(0)-\frac{2\kappa}{v_{0}}}\right]\;.
\end{equation}
The pair annihilation time $t_a$ is determined by $\Delta(t_{a})=0$ and is given by
\begin{equation}
t_{a}=-\frac{\Delta(0)}{v_{0}}-\frac{2\kappa}{v^{2}_{0}}\log\left[1-\frac{v_{0}}{2\kappa}\,\Delta(0)\right]\;.
\end{equation}
For passive systems ($\alpha=0$) this reduces to $t_{a}^0=\Delta^{2}(0)/4\kappa$. This predicts that the annihilation time, normalized to its value in passive systems, $t_{a}/t_{a}^0$, depends only on $\Delta(0)\,v_{0}/2\kappa \sim \alpha\gamma$. Figure ~\ref{fig:trajectories}d shows a fit of the annihilation times extracted from the numerics to this simple formula. The model qualitatively captures the numerical behavior. Note also that defects created a finite distance $\Delta(0)$ apart will always separate provided the activity and hence the magnitude of the self-propulsion velocity exceeds a critical value $ v_0^{c} =2\kappa/\Delta(0)$. 

\section{\label{sec:proliferation}Defect proliferation}

\begin{figure*}[t]
\centering
\includegraphics[width=0.9\textwidth]{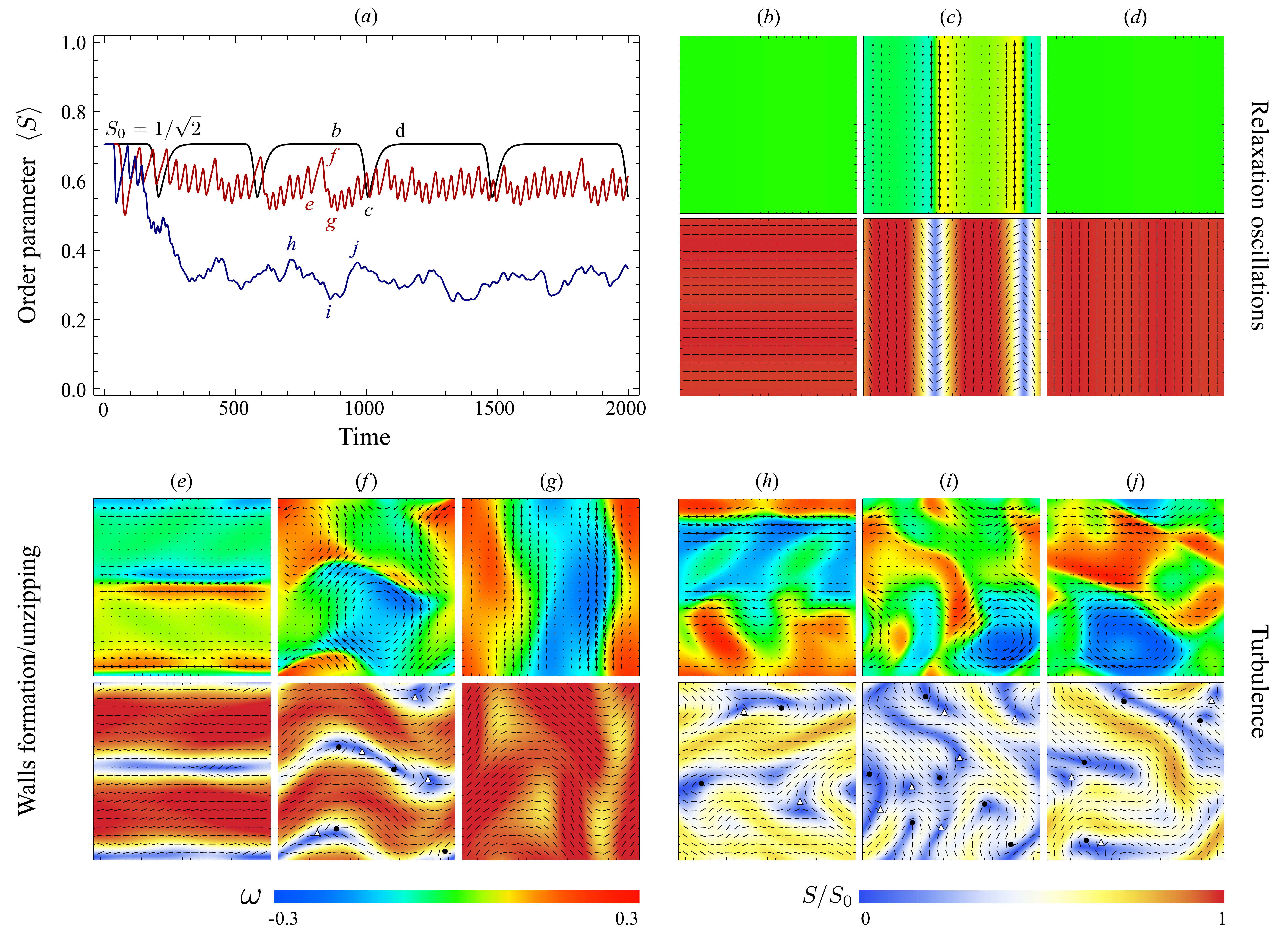}
\caption{\label{fig:zoo}Dynamical states obtained from a numerical integration of Eqs. \eqref{eq:hydrodynamics} with $\gamma=20$ and various values of activity for an extensile system. (a) Average nematic order parameter versus time. The black line for $\alpha=-0.3$ identifies the relaxation oscillations regime with the labels (b), (c) and (d) marking the times corresponding to the snapshots on the top-left panel. The red line for $\alpha=-0.8$ indicates the non-periodic oscillatory regimes characterized by the formation of walls (e) and the unzipping of walls through the unbinding of defect pairs: (f) and (g). The symbols $\bullet$ and $\triangle$ mark the positions of $+1/2$ and $-1/2$ disclinations respectively. The blue line  for $\alpha=-1.2$ corresponds to the turbulent regime in which defects proliferate: (h), (i) and (j). In all the snapshots, the background colors are set by the magnitude of the vorticity $\omega$ and the order parameter $S$ rescaled by the equilibrium value $S_{0}=1/\sqrt{2}$, while the solid lines indicate velocity (top) and director field (bottom). Movies displaying the time evolution of each state are included as the Supplementary Material.}	
\end{figure*}

Pair annihilation is the fundamental mechanism behind defect coarsening in nematic liquid crystals \cite{Bray:2002}. Once this mechanism is suppressed by activity, as described in Sec. \ref{sec:annihilation}, the coarsening dynamics is replaced by a new steady state in which pairs of $\pm 1/2$ disclinations are continuously produced and annihilated at constant rate. The chaotic dynamics following from continuous defect proliferation and annihilation results in turbulent flow \cite{Giomi:2013,Thampi:2013,Thampi:2014a,Thampi:2014b,Gao:2014}. In this section we describe the onset of chaos and the proliferation of defects that are observed from numerical solutions of Eqs. \eqref{eq:hydrodynamics} upon varying the activity parameter $\alpha$ and the rotational viscosity $\gamma$. As initial configurations we  take a homogeneous state with the director field  aligned along the $x-$axis and subject to a small random perturbation in density and orientation. The equations were then integrated from $t = 0$ to $t = 2\times 10^3\tau$ (see Sec. \ref{sec:numerics} for a description of units). 

\begin{figure}[t]
\centering
\includegraphics[width=0.9\columnwidth]{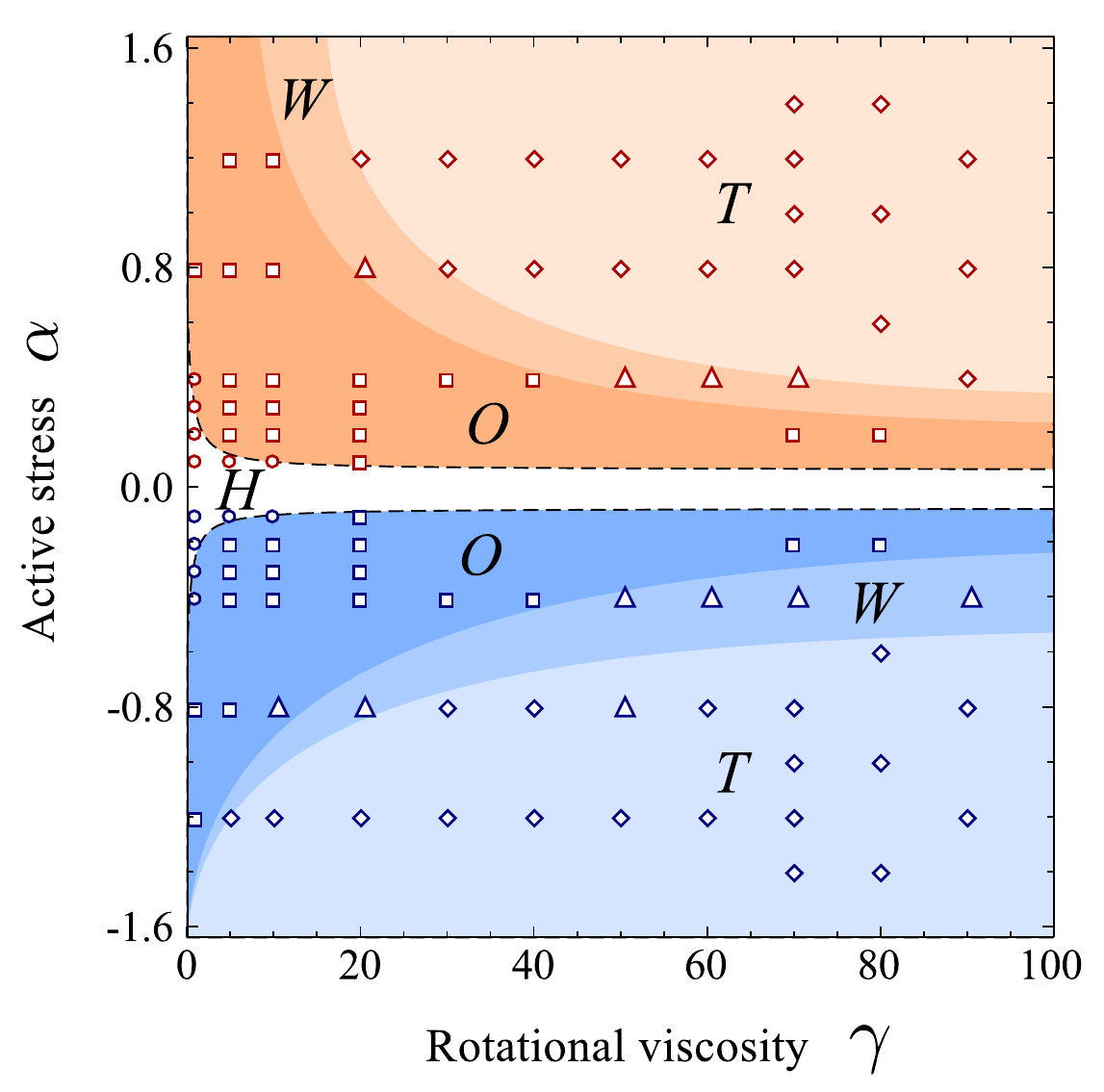}
\caption{Phase diagram showing the various flow regimes of an active nematic obtained by varying activity $\alpha$ and rotational viscosity $\gamma$ for both contractile ($\alpha>0$) and extensile ($\alpha<0$) systems. The dashed lines bounding the region where the homogeneous ordered state ($H$) is stable are the boundaries of linear stability given in  Eq. \eqref{eq:alphac}. With increasing activity, the system exhibits relaxation oscillations ({\em O}), non-periodic oscillations characterized by the formation and unzipping of walls ({\em W}), and turbulence ({\em T}).}
\label{fig:phase_diagram}
\end{figure}

Fig.~\ref{fig:phase_diagram} summarizes the various regimes obtained by exploring the $(\alpha,\gamma)-$plane: {\emph 1)} a homogeneous, quiescent ordered state (H); {\em 2)} a periodic flow marked by the emergence of relaxation oscillations (O); {\em 3)} a non-periodic oscillatory flow characterized by the formation of ``walls'' in the nematic phase and the unzipping of these walls through the unbinding of defect pairs (W); {\em 4)} a chaotic or ``turbulent'' state associated with a constant defect density (T). The latter three regimes are described in more detail in the following.

\subsection{\label{sec:oscillations}Relaxation Oscillations} 

As described in Refs.~\cite{Giomi:2011,Giomi:2012}, relaxation oscillations occur in active nematics when $|\alpha|$ exceeds $\alpha_{2}^{\pm}$ as the result of  the competition of two time scales: the relaxation  $\tau_{\rm p}=\gamma\ell^2/K$ of the nematic structure and the time scale  $\tau_{\rm a}=\eta/|\alpha|$ that controls the rate at which active stresses  are injected in the system. When $\tau_{\rm p}<\tau_{\rm a}$, the microstructure can relax to accommodate the active forcing and the ordered state is stable. This is a quiescent state, with uniform order parameter. Conversely, when $\tau_{\rm p}>\tau_{\rm a}$, the relaxation of the  nematic structure lags behind the injection of active stresses, yielding various dynamical states with spatially and temporally inhomogeneous order parameter. 

In this regime the dynamics consists of a sequence of almost stationary passive periods separated by active ``bursts'' in which the director switches abruptly between two orthogonal orientations (Fig. \ref{fig:zoo}b-d). During passive periods, the particle concentration and the nematic order parameter are nearly uniform across the system, there is no appreciable flow, and the director field is either parallel or perpendicular to the $x$ direction (due to the initial conditions). Eventually this configuration breaks down and the director field rotates by 90$^{\circ}$. The rotation of the director field is initially localized along narrow extended regions, generating flowing bands similar to those obtained in active nematic films \cite{Voituriez:2005} (Fig. \ref{fig:zoo}c). The temporary distortion of the director field as well as the formation of the bands is accompanied by the onset of flow along the longitudinal direction of the bands, with neighboring bands flowing in opposite directions. The flow terminates after the director field rotates and a uniform orientation is restored. The process then repeats. 

Depending on whether the active stress fueling the oscillatory dynamics is contractile or extensile, the rotation of the director occurs through an intermediate splay or bending deformation. During bursts, the nematic order parameter, otherwise equal to its equilibrium value, drops significantly (Fig. \ref{fig:zoo}a, black line). Without this transient melting the distortions of the director field required for a burst are unfavorable for any level of activity.

The frequency of the oscillation is proportional to $k^{2}\alpha$, where $k=2\pi/L$ is the wave number of the longest-wavelength mode to go unstable \cite{Giomi:2011,Giomi:2012}. In spite of the strong elastic deformation and the dramatic drop in the order parameter, this regime contains no unbound defects.

\subsection{\label{sec:walls}Wall formation and unzipping} 

For larger values of activity the bend and splay deformations of the director at the band boundaries become large enough to drive creation  of defect pairs, as shown in Fig. \ref{fig:zoo}e-g. The alignment of the bands again oscillates between two orthogonal directions ($x-$ and $y-$axis for the initial condition used here), but the switching takes place through an intermediate more complex configuration with lozenge-shaped ordered regions. Defect pairs then unbind and glide along the narrow regions separating two bands where elastic deformations and shear flows are largest. These regions of high distortion and shear rate are commonly referred to as ``walls'' in the liquid crystals literature~\cite{Helfrich:1968}. Movies displaying the dynamics of wall unzipping and defect creation for both extensile and contractile systems are included as Supplementary Material.

\begin{figure}[t]
\centering
\includegraphics[width=0.8\columnwidth]{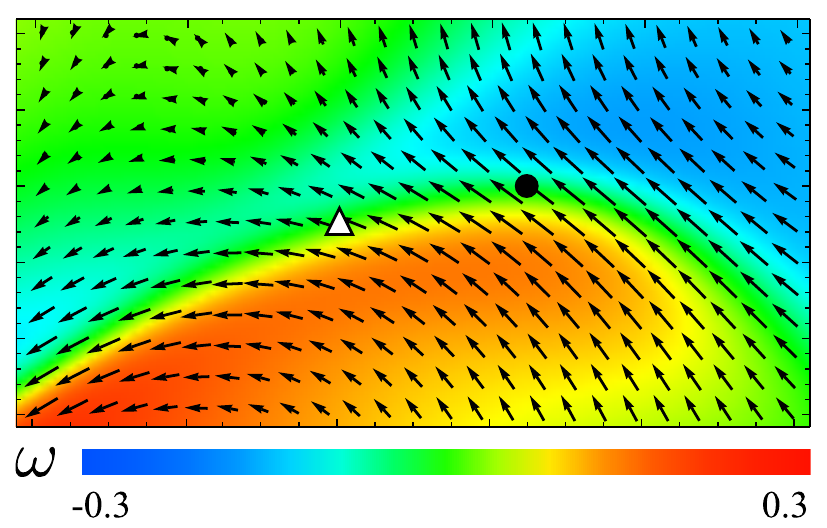}
\caption{\label{fig:defect-creation} A magnification of the snapshot of Fig. \ref{fig:zoo}f, showing the creation of a $+1/2$ ($\bullet$) and a $-1/2$ ($\triangle$) defect pair along a wall or $\alpha=-0.8$ and $\gamma=20$. The black arrows indicate the flow velocity, while the background color is related with the local vorticity. The wall is also the boundary between a pair of vortices of opposite circulation. The flow field of opposite-signed vortices adds at the wall, yielding a region of high shear that promotes defect unbinding.}
\end{figure}

The formation of walls and the ``unzipping'' of defects along the walls by the creation of pairs of $\pm 1/2$ disclinations has being discussed in detail by Thampi {\em et al}. \cite{Thampi:2014a} (see also Ref. \cite{Thampi:2014b} in this Themed Issue). Defect unbinding along the walls relaxes both the excess elastic energy and the high shear stresses present in these regions, where the nematic order parameter $S$ is driven near zero. For this reason wall formation and unzipping occurs often at the boundary between pairs of vortices of opposite circulation (Fig. \ref{fig:defect-creation}). In passive nematic liquid crystals the strong bending deformation that leads to the formation of walls, and preceedes defect unbinding, requires an external action, such as an applied electric field or an externally imposed shear stress \cite{Lozar:2005}. In active nematics, on the other hand, a similar complex spatiotemporal dynamics occurs spontaneously, driven by the \emph{local} injection of active stresses, which are in turn balanced by spontaneous distortions and flows as described in Section \ref{sec:linearstability}. Wall formation and unzipping also combines here with the oscillatory dynamics described in Section \ref{sec:oscillations} to give rise to the periodic creation and annihilation of topological defects that marks the transition to the turbulent regime.

Oscillating band structures of the type observed here are found in passive nematic fluids under externally applied shear flows and are precursors to rheochaos~\cite{Chakrabarti:2004,Chakraborty:2010}. They have been predicted theoretically~\cite{Fielding:2004} and  observed experimentally~\cite{Ganapathy:2008} in suspensions of wormlike micelles, where the spatiotemporal dynamics is directly correlated to shear banding~\cite{Olmsted:2008} and to stress/shear fluctuations at the shear band interface.  In passive systems it has been argued that the route to rheochaos depends on whether the stress or the strain rate is controlled during the experiment.  It would be interesting to analyze in more detail the route to chaos in this case.

\subsection{\label{sec:turbulence}Turbulence}

\begin{figure}[t]
\centering
\includegraphics[width=0.9\columnwidth]{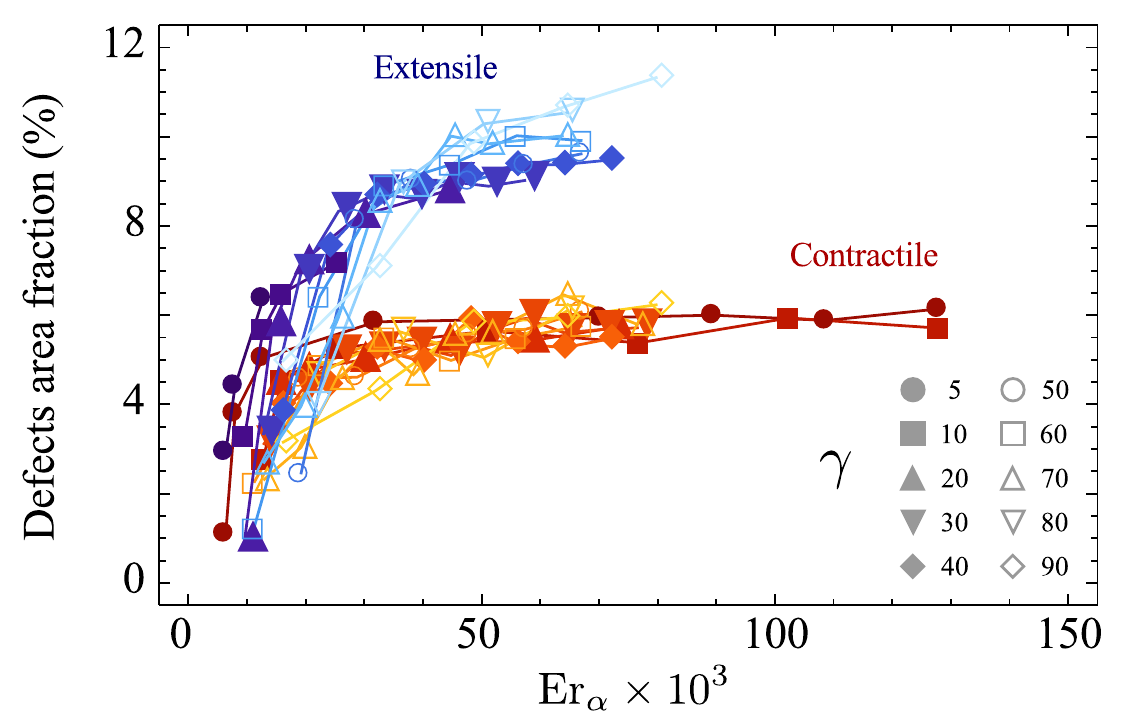}
\caption{\label{fig:defect-scaling}Defects area fraction $N\pi a^{2}/L^{2}$ as a function of the active Ericksen number ${\rm Er}_{\alpha}=\alpha\gamma L^{2}/(\eta K)$ for contractile (symbols in red tones) and extensile (symbols in blue tones) systems. In both cases the area fraction saturates when the activity increase is compensated by a drop of the order parameter which effectively reduces the injected active stress.}
\end{figure}

At even higher activity the bands/walls structure begins to bend  and  fold and the dynamics becomes chaotic, resembling that of a ``turbulent'' fluid, as displayed in Fig. \ref{fig:zoo}h-j and in the movies included as Supplementary Material. The system reaches a dynamical steady state where defect pairs are continuously created and annihilated, but their mean number remains on average constant in time. Due to topological charge conservation, at any time the system contains an equal number of positive and negative defects. Unlike in equilibrium systems, however, in active nematics it is possible for opposite-charge defects of certain orientations to repel instead of attracting (see Sec. \ref{sec:annihilation}). This allows the formation of a defective steady state, with self-sustained flows and a constant mean number of defects. 

The defective dynamics observed in the system is similar to that obtained in a passive nematic subject to an externally imposed shear or to electrohydrodynamic instabilities \cite{Manneville:1981}.  In active nematics, however, defects are generated in the absence of any externally imposed forces or constraints, as a result of the spontaneous distortion induced by the active stresses. After formation the defects are convected by the swirling flow and interact with one another through distortional elasticity as well as hydrodynamically through the modifications the defects themselves induce on the flow field. 

The sequence of flow regimes observed here is reminiscent of the evolution of the dynamics of sheared tumbling nematic polymers with increasing shear rate, known as the Ericksen number cascade~\cite{Larson:1993}. In a nematic film sheared at a rate $\dot\epsilon$ between two plates separated by a distance $h$, the Ericksen number ${\rm Er}$ provides a dimensionless  measure of the magnitude of viscous torques  ($\sim\dot\epsilon$) relative to elastic torques arising from spatial gradients in the average molecular orientation ($\sim K/\gamma h^{2})$, with  ${\rm Er}=\dot\epsilon\gamma h^2/K$. This suggest the definition of an \emph{active} Ericksen number ${\rm Er}_{\alpha}$ where the active stress $\alpha$ takes the place of viscous stresses $\eta\dot{\epsilon}$. The active Ericksen number is then defined as ${\rm Er}_{\alpha}=\alpha \gamma L^2/(\eta K)$. With this definition, ${\rm Er}_{\alpha}$ can also be interpreted as the ratio of the time scale $\tau_{\rm p}=\gamma L^2/K$ for the relaxation of an orientational deformation of the nematic order on the scale of the entire system to the time $\tau_{\rm a}=\eta/\alpha$ controlling the injection of active stresses.

Fig. \ref{fig:defect-scaling} shows a plot of the {\em area fraction} occupied by defects as a function of the active Ericksen number ${\rm Er}_{\alpha}$ for both extensile and contractile systems. The area fraction is defined as the relative area occupied by the core of the defects, or $N\pi a^{2}/L^{2}$, with $N$ the number of defects. The core radius $a$ resulting from the hydrodynamic equations \eqref{eq:hydrodynamics} is approximatively given by the size of the boundary layer between the position of a defect, where $S=0$, and surrounding space, where $S=S_{0}$. This is proportional to the coefficient $A$ in the Landau-de Gennes free energy \eqref{eq:free_energy}
\begin{equation}\label{eq:core_radius}
a = \sqrt{\frac{K}{|A|}} \approx \frac{1}{\sqrt{|c_{0}-c^{*}|}} \approx \ell\;.
\end{equation}
For larger activity the reduction of the nematic order parameter due to the unbound defects compensates the activity increase by effectively reducing the injected local stress $\sigma^{\rm a}\approx \alpha S$.

For the same magnitude of activity, extensile systems contain a larger mean number of defects than contractile ones. In both case defect pairs first unbind within the walls, which are regions of large bend and splay deformations in extensile and contractile systems, respectively. As the system  evolves toward the regime of chaotic dynamics, the walls begin to deform largely via bend deformations in both systems \cite{Thampi:2014a,Thampi:2014b}. This asymmetry could be because the severe splay deformations localized at the walls lead to a more drastic reduction of the nematic order parameter. Thus contractile flow-tumbling nematics, whose spontaneous distortion involves mostly splay deformations, are effectively less active than flow-tumbling extensile systems.

\section{\label{sec:discussion}Discussion and Conclusions}
 
We have presented a detailed analytical and numerical study of the mechanics of topological defects in active nematic liquid crystals. Topological defects are distinctive signatures of liquid crystals and profoundly affect their viscoelastic behavior by constraining the orientational structure of the fluid in a way that inevitably requires system-wide (global) changes not achievable with any set of local deformations. In ordered states of both passive and active nematics, the topological defects are fingerprints of the broken symmetry in the ordered state. In particular, the presence of strength $\pm 1/2$ defects clearly reveals the nematic nature of the orientational order, in contrast to systems with polar (ferromagnetic) symmetry where the lowest energy defects allowed have strength $\pm1$.  Active liquid crystals have the additional feature that defects act as local sources of motion, behaving as  self-propelled particle-like objects (see Sec. \ref{sec:isolated}). The direction of motion of the strength $+1/2$ defects provides, furthermore, a clear signature of the extensile or contractile nature of the active stresses, as the comet-like positive defects are advected towards their head in extensile systems and towards their tail in contractile ones. In passive liquid crystals defect dynamics is always transient, as oppositely charged defects attract and eventually annihilate. In active nematics, on the other hand, the interplay between active and viscous stresses, modulated by the director geometry induced by the defects, enriches the spectrum of defect-defect interactions by allowing for an effective repulsion between defects and anti-defects (see Sec. \ref{sec:annihilation}). For highly active systems this mechanism can arrest the process of coarsening, leading to a state where unbound pairs of defects are continuously created and annihilated, but with their mean density constant in time (see Sec. \ref{sec:proliferation}). The chaotic dynamics originating from the continuous defect proliferation and annihilation results in spontaneous low Reynolds number turbulence, akin to the so-called director turbulence seen in sheared polymer and micellar nematics~\cite{Larson:1993}.

Several open questions remain concerning the defect dynamics of active nematics. The defect area fraction shown in Fig.~\ref{fig:defect-scaling}  exhibits a crossover from growth at low activity to saturation at high activity, but an understanding of the length scales that control this behavior is still lacking. Both our work and work by Thampi {\em et al}.~\cite{Thampi:2014b} suggest that the mean separation between defects in the turbulent regime coincides with the typical vortex size, but more work is needed to elucidate the behavior of this length scale with activity over a wide range of parameters. While defect generation in sheared nematics has been explained in terms of a simple rate equation that balances creation and annihilation~\cite{Larson:1993}, a similar simple model for active defects is still missing. On the basis of numerics, Thampi {\em et al}. have suggested that the defect creation rate should scale like the square of the activity~\cite{Thampi:2014b}, but no simple argument is available to understand this counterintuitive result. Finally, an important open question is the different behavior of extensile and contractile systems apparent from Fig.~\ref{fig:defect-scaling}. The turbulent state of contractile systems is much less defective than that of extensile ones, suggesting that for equal magnitude of the activity $\alpha$, contractile nematics are effectively ``less active'' than extensile ones. 

\acknowledgements

We thank Zvonimir Dogi{\'c}, Suzanne Fielding, Jean-Francois Joanny, Oleg Lavrentovich, and Tim Sanchez  for several illuminating discussions. LG was supported by SISSA mathLab. The work at Syracuse was supported by the National Science Foundation through awards DMR-1305184 (MCM, PM) and DGE-1068780 (MCM) and by funds from the Syracuse Soft Matter Program. MCM also acknowledges support from the Simons Foundation. Finally, MCM and MJB thank  the KITP at the University of California, Santa Barbara, for support during the preparation of the manuscript.

\appendix

\begin{widetext}
\section{Active backflow of $+1/2$ disclinations}

Finding the backflow produced by the positive disclination reduces to the calculation of the following integrals:
\begin{subequations}\label{eq:integrals}
\begin{gather}
I_{1} = \int dA'\,\frac{1}{r'}\left(\log\frac{\mathcal{L}}{|\bm{r}-\bm{r}'|}-1\right)\;,\\[10pt]
I_{2} = \int dA'\,\frac{1}{r'}\frac{(r_{i}-r'_{i})(x-x')}{|\bm{r}-\bm{r}'|^{2}}\;.
\end{gather}	
\end{subequations}
To calculate the first integral, we can make use of the logarithmic expansion
\begin{equation}\label{eq:log_expansion}
\log \frac{|\bm{r}-\bm{r}'|}{\mathcal{L}} = \log\left(\frac{r_{>}}{\mathcal{L}}\right)-\sum_{n=1}^{\infty} \frac{1}{n}\left(\frac{r_{<}}{r_{>}}\right)^{n}\cos [n(\phi-\phi')]\;,	
\end{equation}
where $r_{>}=\max (|\bm{r}|,|\bm{r}'|)$ and $r_{<}=\min (|\bm{r}|,|\bm{r}'|)$. The integral over the angle can be immediately carried out using the orthogonality of trigonometric functions: $\int_{0}^{2\pi} d\phi'\,\cos [n(\phi-\phi')] = \delta_{n0}$. The remaining radial integral can be then straightforwardly calculated:
\begin{equation}\label{eq:step_1}
\int dr'\,\log\left(\frac{r_{>}}{\mathcal{L}}\right) = r-R+R\log\left(\frac{R}{\mathcal{L}}\right)\;.
\end{equation}	
Thus:
\[
I_{1} = -2\pi \left[r+R\log\left(\frac{R}{\mathcal{L}}\right)\right]\,.
\] 
To calculate the integral $I_{2}$, we can make use of the fact that
\begin{equation}\label{eq:trick}
\frac{r_{i}-r'_{i}}{|\bm{r}-\bm{r}'|^{2}} = \frac{\partial}{\partial r_{i}}\log |\bm{r}-\bm{r}'|\;.
\end{equation}
Thus one can write
\begin{equation}\label{eq:i2}
I_{2} = x\,\frac{\partial}{\partial r_{i}}\int dr'\,d\phi'\,\log |\bm{r}-\bm{r'}|-\frac{\partial}{\partial r_{i}}\int dr'\,d\phi'\,x'\log |\bm{r}-\bm{r'}|\;.
\end{equation}
The first integral is calculated by differentiating $I_{1}$:
\begin{equation}\label{eq:i2a}
x\frac{\partial}{\partial r_{i}}\int dr'\,d\phi\,\log|\bm{r}-\bm{r}'| = \frac{2\pi x r_{i}}{r}\;.
\end{equation}
The second integral in Eq. \eqref{eq:i2} can be calculated again with the help of the logarithmic expansion \eqref{eq:log_expansion}:
\begin{equation}
\int dr'\,d\phi'\,r\cos\phi'\log |\bm{r}-\bm{r}'| 
= -\pi \cos\phi \int_{0}^{R} dr'\,r'\,\frac{r_{<}}{r_{>}} 
= -\pi r\cos\phi \left(R-\frac{2}{3}\,r\right)\;,
\end{equation}
where we have used again the orthogonality of trigonometric functions: 
\begin{equation}\label{eq:orthogonality}
\int_{0}^{2\pi}d\phi' \cos n(\phi-\phi')\cos m\phi' = \pi \cos m\phi\,\delta_{nm}\;.
\end{equation}
Thus, taking the derivative and combining with Eq. \eqref{eq:i2a}, yields:
\begin{equation}
I_{2} = \frac{4\pi}{3}\,\frac{x r_{i}}{r} +  \pi \left(R-\frac{2}{3}\,r\right)\delta_{ix}\;.
\end{equation}
Adding together $I_{1}$ and $I_{2}$ and switching to polar coordinates gives:
\begin{subequations}\label{eq:solution_before_simplifications}
\begin{gather}
v_{x} =\frac{\alpha}{12\eta}\left\{\left[3\left(\frac{3R}{2}+3R\log\frac{\mathcal{L}}{R}-r\right)+r\cos 2\phi\right]\right\}\;,\\
v_{y} =\frac{\alpha}{12\eta}\,r\sin 2\phi\;.
\end{gather}
\end{subequations}
Setting $\mathcal{L}=R\sqrt{e}$ and expressing everything in polar coordinates one finally gets Eq. (\ref{eq:active_backflow}a).

\section{Active backflow of $-1/2$ disclinations}

The calculation of the active backflow associated with a negative disclination reduces to the calculation of the following integrals:
\begin{gather}
I_{3} = \int dA'\,\frac{1}{r'}\left(\log\frac{\mathcal{L}}{|\bm{r}-\bm{r}'|}-1\right)\,\left[\sin 2\phi'\,\bm{\hat{y}}-\cos 2\phi'\,\bm{\hat{x}}\right]\;,\\[10pt]
I_{4} = \int dA'\,\frac{1}{r'}\frac{r_{i}-r_{i}'}{|\bm{r}-\bm{r}'|^{2}}\,\left[(y-y')\sin 2\phi'-(x-x')\cos 2\phi'\right].
\end{gather}
The first integral can be calculated with the help of the logarithmic expansion \eqref{eq:log_expansion} as well as the orthogonality condition \eqref{eq:orthogonality}. This yields:
\begin{equation}
I_{3} = \frac{1}{2}\,\pi r \left(\frac{4}{3}-\frac{r}{R}\right)(\sin 2\phi\,\bm{\hat{y}}-\cos 2\phi\bm{\hat{x}})\;.
\end{equation}
To calculate $I_{4}$ we can use again Eq. \eqref{eq:trick}. Thus:
\begin{multline}\label{eq:long_equation}
I_{4} 
= y \frac{\partial}{\partial r_{i}}\int dr' d\phi'\,\log|\bm{r}-\bm{r}'|\,\sin 2\phi'
- \frac{\partial}{\partial r_{i}}\int dr' d\phi'\,\log|\bm{r}-\bm{r}'|\,y'\sin 2\phi' \\
- x \frac{\partial}{\partial r_{i}}\int dr' d\phi'\,\log|\bm{r}-\bm{r}'|\,\cos 2\phi'
+ \frac{\partial}{\partial r_{i}}\int dr' d\phi'\,\log|\bm{r}-\bm{r}'|\,x'\cos 2\phi'\;.
\end{multline}
The first and third integral in Eq. \eqref{eq:long_equation} are respectively the opposite of the $y-$ and $x-$component of $I_{3}$. The remaining two integrals, can be calculated using Eq. \eqref{eq:log_expansion} and \eqref{eq:orthogonality}. This gives:
\begin{equation}
\int dr' d\phi'\, \log |\bm{r}-\bm{r}'|\,f_{\pm}(r',\phi') = -\frac{\pi}{2} \left[r\left(R-\frac{2}{3}r\right)\cos\phi\pm\frac{1}{3}r^{2}\left(\frac{6}{5}-\frac{r}{R}\right)\cos 3\phi\right]\;,
\end{equation}
where $f_{+}(r',\phi')=x'\cos 2\phi'$ and $f_{-}(r',\phi')=y'\sin 2\phi'$. Combining $I_{3}$ and $I_{4}$ and switching to polar coordinates finally gives Eq. (\ref{eq:active_backflow}b).
\end{widetext}

\end{document}